\documentclass[a4paper,showpacs,showkeys,nofootinbib,aps]{revtex4}
\usepackage{epsfig}
\begin{document}
\title{Azimuthal Asymmetries in DIS as a Probe \\
of Intrinsic Charm Content of the Proton}
\author{L.N.~Ananikyan}
 \email{lev@web.am}
 \affiliation{Yerevan Physics Institute, Alikhanian Br.2, 375036 Yerevan, Armenia}
\author{N.Ya.~Ivanov}
 \email{nikiv@uniphi.yerphi.am}
\affiliation{Yerevan Physics Institute, Alikhanian Br.2, 375036 Yerevan, Armenia}
\date{\today}
\begin{abstract}
\noindent We calculate the azimuthal dependence of the heavy-quark-initiated ${\cal O}(\alpha_{s})$
contributions to the lepton-nucleon deep inelastic scattering (DIS). It is shown that, contrary to
the photon-gluon fusion (GF) component, the photon-quark scattering (QS) mechanism is practically
$\cos2\varphi$-independent. We investigate the possibility to discriminate experimentally between
the GF and QS contributions using their strongly different azimuthal distributions. Our analysis
shows that the GF and QS predictions for the azimuthal $\cos2\varphi$ asymmetry are quantitatively
well defined in the fixed flavor number scheme: they are stable, both parametrically and
perturbatively. We conclude that measurements of the azimuthal distributions at large Bjorken $x$
could directly probe the intrinsic charm content of the proton. As to the variable flavor number
schemes, the charm densities of the recent CTEQ and MRST sets of parton distributions have a
dramatic impact on the $\cos2\varphi$ asymmetry in the whole region of $x$ and, for this reason,
can easily be measured.
\end{abstract}
\maketitle
\section{Introduction}
The notion of the intrinsic charm (IC) content of the proton has been introduced over 25 years ago
in Refs~\cite{BHPS,BPS}. It was shown that, in the light-cone Fock space picture
\cite{brod1,brod2}, it is natural to expect a five-quark state contribution to the proton wave
function. The probability to find in a nucleon the five-quark component $\left\vert
uudc\bar{c}\right\rangle$ is of higher twist since it scales as $1/m^{2}$ where $m$ is the
$c$-quark mass \cite{polyakov}. This component can be generated by $gg\rightarrow c\bar{c}$
fluctuations inside the proton where the gluons are coupled to different valence quarks. Since all
of the quarks tend to travel coherently at same rapidity in the $\left\vert
uudc\bar{c}\right\rangle $ bound state, the heaviest constituents carry the largest momentum
fraction. For this reason, one would expect that the intrinsic charm component to be dominate the
$c$ -quark production cross sections at sufficiently large Bjorken $x$. So, the original concept of
the charm density in the proton \cite{BHPS,BPS} has nonperturbative nature and will be referred to
in the present paper as nonperturbative IC.

A decade ago another point of view on the charm content of the proton has been proposed in the
framework of the variable flavor number scheme (VFNS) \cite{ACOT,collins}. The VFNS is an approach
alternative to the traditional fixed flavor number scheme (FFNS) where only light degrees of
freedom ($u,d,s$ and $g$) are considered as active. It is well known that a heavy quark production
cross section contains potentially large logarithms of the type $\alpha_{s}\ln\left(
Q^{2}/m^{2}\right)$ whose contribution dominates at high energies, $Q^{2}\rightarrow\infty$. Within
the VFNS, these mass logarithms are resummed through the all orders into a heavy quark density
which evolves with $Q^{2}$ according to the standard DGLAP \cite{grib-lip,dokshitzer,alt-par}
evolution equation. Hence the VFN schemes introduce the parton distribution functions (PDFs) for
the heavy quarks and change the number of active flavors by one unit when a heavy quark threshold
is crossed. We can say that the charm density arises within the VFNS perturbatively via the
$g\rightarrow c\bar{c}$ evolution and will call it the perturbative IC.

Presently, both perturbative and nonperturbative IC are widely used for a phenomenological
description of available data. (A recent review of the theory and experimental constraints on the
charm quark distribution can be found in Refs.~\cite{pumplin,brod-higgs}. See also
Appendix~\ref{exp} in the present paper). In particular, practically all the recent versions of the
CTEQ \cite{CTEQ6} and MRST \cite{MRST2004} sets of PDFs are based on the VFN schemes and contain a
charm density. At the same time, the key question remains open: How to measure the intrinsic charm
content of the proton? The basic theoretical problem is that radiative corrections to the fixed
order predictions for the production cross sections are large. In particular, the next-to-leading
order (NLO) corrections increase the leading order (LO) results for both charm and bottom
production cross sections by approximately a factor of two at energies of the fixed target
experiments. Moreover, soft gluon resummation of the threshold Sudakov logarithms indicates that
higher-order contributions are also essential. (For a review see Refs.~\cite{kid2,kid1}). On the
other hand, perturbative instability leads to a high sensitivity of the theoretical calculations to
standard uncertainties in the input QCD parameters. For this reason, it is difficult to compare
pQCD results for spin-averaged cross sections with experimental data directly, without additional
assumptions. The total uncertainties associated with the unknown values of the heavy quark mass,
$m$, the factorization and renormalization scales, $\mu _{F}$ and $\mu _{R}$, $ \Lambda _{QCD}$ and
the PDFs are so large that one can only estimate the order of magnitude of the pQCD predictions for
production cross sections \cite{Mangano-N-R,Frixione-M-N-R}.

Since production cross sections are not perturbatively stable, it is of special interest to study
those observables that are well-defined in pQCD. A nontrivial example of such an observable was
proposed in Refs.~\cite{we1,we2,we4,we3} where the azimuthal $\cos2\varphi$ asymmetry in heavy
quark photo- and leptoproduction has been analyzed~\footnote{The well-known examples are the shapes
of differential cross sections of heavy flavor production which are sufficiently stable under
radiative corrections.}. In particular, the Born level results have been considered \cite{we1} and
the NLO soft-gluon corrections to the basic mechanism, photon-gluon fusion (GF), have been
calculated \cite{we2,we4}. It was shown that, contrary to the production cross sections, the
azimuthal asymmetry in heavy flavor photo- and leptoproduction is quantitatively well defined in
pQCD: the contribution of the dominant GF mechanism to the asymmetry is stable, both parametrically
and perturbatively. Therefore, measurements of this asymmetry would provide an ideal test of pQCD.
As was shown in Ref.~\cite{we3}, the azimuthal asymmetry in open charm photoproduction could have
been measured with an accuracy of about ten percent in the approved E160/E161 experiments at SLAC
\cite{E161} using the inclusive spectra of secondary (decay) leptons.

In the present paper we study the IC contribution to the azimuthal asymmetries in heavy quark
leptoproduction:
\begin{equation}
l(\ell )+N(p)\rightarrow l(\ell -q)+Q(p_{Q})+X[\overline{Q}](p_{X}). \label{1}
\end{equation}
Neglecting the contribution of $Z-$boson as well as the target mass effects, the cross section of
the reaction (\ref{1}) for unpolarized initial states  may be written as
\begin{equation}\label{2}
\frac{\text{d}^{3}\sigma _{lN}}{\text{d}x\text{d}Q^{2}\text{d}\varphi }=\frac{\alpha _{em}}{(2\pi
)^{2}}\frac{1}{xQ^{2}}\frac{y^2}{1-\varepsilon}\left[\sigma _{T}( x,Q^{2})+ \varepsilon\sigma_{L}(
x,Q^{2})+ \varepsilon\sigma_{A}( x,Q^{2})\cos 2\varphi+
2\sqrt{\varepsilon(1+\varepsilon)}\sigma_{I}( x,Q^{2})\cos \varphi\right],
\end{equation}
The quantity $\varepsilon$ measures the degree of the longitudinal polarization of the virtual
photon in the Breit frame \cite{dombey},
\begin{equation}\label{3}
\varepsilon=\frac{2(1-y)}{1+(1-y)^2},
\end{equation}
and the kinematic variables are defined by
\begin{eqnarray}
\bar{S}=\left( \ell +p\right) ^{2},\qquad &Q^{2}=-q^{2},\qquad &x=\frac{Q^{2}}{%
2p\cdot q},  \nonumber \\
y=\frac{p\cdot q}{p\cdot \ell },\qquad &Q^{2}=xy\bar{S},\qquad &\rho =\frac{4m^{2}%
}{\bar{S}}.  \label{4}
\end{eqnarray}
The cross sections $\sigma _{i}$ $(i=T,L,A,I)$ in Eq.~(\ref{2}) are related to the structure
functions $F_{i}(x,Q^{2})$ as follows:
\begin{eqnarray}
F_{i}(x,Q^{2}) &=&\frac{Q^{2}}{8\pi^{2}\alpha _{em}x}\,\sigma_{i}(x,Q^{2}), \qquad (i=T,L,A,I)\nonumber \\
F_{2}(x,Q^{2}) &=&\frac{Q^{2}}{4\pi^{2}\alpha _{em}}\,\sigma_{2}(x,Q^{2}),\label{6}
\end{eqnarray}
where $F_{2}=2x(F_{T}+F_{L})$ and $\sigma_{2}=\sigma_{T}+\sigma_{L}$.
\begin{figure}
\begin{center}
\mbox{\epsfig{file=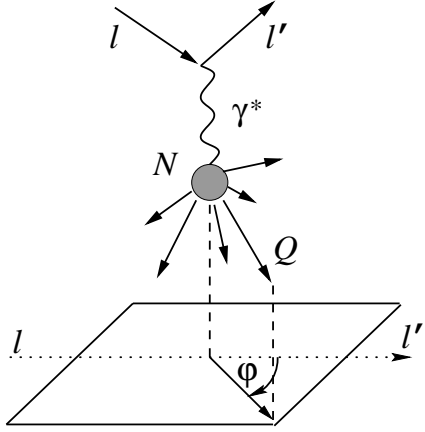,width=200pt}}
\caption{\label{Fg.1}\small Definition of the azimuthal
angle $\varphi$ in the nucleon rest frame.}
\end{center}
\end{figure}
In Eq.~(\ref{2}), $\sigma _{T}\,(\sigma _{L})$ is the usual $\gamma ^{*}N$ cross section describing
heavy quark production by a transverse (longitudinal) virtual photon. The third cross section,
$\sigma _{A}$, comes about from interference between transverse states and is responsible for the
$\cos2\varphi$ asymmetry which occurs in real photoproduction using linearly polarized photons
\cite{we1,we2,we3}. The fourth cross section, $\sigma _{I}$, originates from interference between
longitudinal and transverse components \cite{dombey}. In the nucleon rest frame, the azimuth
$\varphi $ is the angle between the lepton scattering plane and the heavy quark production plane,
defined by the exchanged photon and the detected quark $Q$ (see Fig.~\ref{Fg.1}). The covariant
definition of $\varphi $ is
\begin{eqnarray}
\cos \varphi &=&\frac{r\cdot n}{\sqrt{-r^{2}}\sqrt{-n^{2}}},\qquad \qquad
\sin \varphi =\frac{Q^{2}\sqrt{1/x^{2}+4m_{N}^{2}/Q^{2}}}{2\sqrt{-r^{2}}%
\sqrt{-n^{2}}}~n\cdot \ell ,  \label{7} \\
r^{\mu } &=&\varepsilon ^{\mu \nu \alpha \beta }p_{\nu }q_{\alpha }\ell _{\beta },\qquad \qquad
\quad n^{\mu }=\varepsilon ^{\mu \nu \alpha \beta }q_{\nu }p_{\alpha }p_{Q\beta }.  \label{8}
\end{eqnarray}
In Eqs.~(\ref{4}) and (\ref{7}), $m$ and $m_{N}$ are the masses of the heavy quark and the target,
respectively. Usually, the azimuthal asymmetry associated with the $\cos 2\varphi $ distribution,
$A_{2\varphi}(\rho ,x,Q^{2})$, is defined by
\begin{eqnarray}
A_{2\varphi}(\rho ,x,Q^{2})&=&2\langle \cos 2\varphi \rangle(\rho ,x,Q^{2})=\frac{\text{d}^{3}
\sigma _{lN}(\varphi =0)+\text{d}^{3}\sigma _{lN}(\varphi =\pi )-2\text{d}^{3}\sigma _{lN}(\varphi
=\pi /2)}{ \text{d}^{3}\sigma _{lN}(\varphi =0)+\text{d}^{3}\sigma _{lN}(\varphi =\pi
)+2\text{d}^{3}\sigma _{lN}(\varphi =\pi /2)} \nonumber\\
&=&\frac{\varepsilon \,\sigma _{A}( x,Q^{2}) }{\sigma _{T}( x,Q^{2}) +\varepsilon \,\sigma _{L}(
x,Q^{2}) }=A(x,Q^{2})\frac{\varepsilon+\varepsilon R(x,Q^{2})}{1+\varepsilon R(x,Q^{2})},
\label{9}
\end{eqnarray}
where $\text{d}^{3}\sigma _{lN}(\varphi )\equiv {{\displaystyle{\text{d}^{3}\sigma _{lN} \over
\text{d}x\text{d}Q^{2}\text{d}\varphi }} }( \rho ,x,Q^{2},\varphi)$ and the mean value of $\cos
n\varphi$ is
\begin{equation}
\langle \cos n\varphi \rangle (\rho ,x,Q^{2})= \frac{\int\limits_{0}^{2\pi }\text{d}\varphi \cos
n\varphi {\displaystyle {\text{d}^{3}\sigma _{lN} \over \text{d}x\text{d}Q^{2}\text{d}\varphi }} (
\rho ,x,Q^{2},\varphi ) }{\int\limits_{0}^{2\pi }\text{d}\varphi {\displaystyle {\text{d}^{3}\sigma
_{lN} \over \text{d}x\text{d}Q^{2}\text{d}\varphi }} ( \rho ,x,Q^{2},\varphi ) }.  \label{10}
\end{equation}
In Eq.~(\ref{9}), the quantities $R(x,Q^{2})$ and $A(x,Q^{2})$ are defined as
\begin{eqnarray}
R(x,Q^{2})&=&\frac{\sigma_{L}}{\sigma_{T}}(x,Q^{2})=\frac{F_{L}}{F_{T}}(x,Q^{2}), \label{11}\\
A(x,Q^{2})&=&\frac{\sigma_{A}}{\sigma_{2}}(x,Q^{2})=2x\frac{F_{A}}{F_{2}}(x,Q^{2}). \label{12}
\end{eqnarray}
Likewise, we can define the azimuthal asymmetry associated with the $\cos \varphi $ distribution,
$A_{\varphi}(\rho ,x,Q^{2})$:
\begin{eqnarray}
A_{\varphi}(\rho ,x,Q^{2})&=&2\langle \cos \varphi \rangle(\rho ,x,Q^{2})=\frac{2\text{d}^{3}
\sigma _{lN}(\varphi =0)-2\text{d}^{3}\sigma _{lN}(\varphi =\pi )}{ \text{d}^{3}\sigma
_{lN}(\varphi =0)+\text{d}^{3}\sigma _{lN}(\varphi =\pi
)+2\text{d}^{3}\sigma _{lN}(\varphi =\pi /2)} \nonumber\\
&=&\frac{2\sqrt{\varepsilon(1+\varepsilon)}\,\sigma _{I}( x,Q^{2}) }{\sigma _{T}( x,Q^{2})
+\varepsilon\,\sigma_{L}(x,Q^{2})}=A_{I}(x,Q^{2})\sqrt{\varepsilon(1+\varepsilon)/2}
\frac{1+R(x,Q^{2})}{1+\varepsilon R(x,Q^{2})}, \label{13}
\end{eqnarray}
where
\begin{equation}\label{14}
A_{I}(x,Q^{2})=2\sqrt{2}\frac{\frac{}{}\sigma_{I}}{\sigma_{2}}(x,Q^{2})=
4\sqrt{2}\,x\frac{F_{I}}{F_{2}}(x,Q^{2}).
\end{equation}
Remember that $y\ll 1$ in most of the experimentally reachable kinematic range. Taking also into
account that $\varepsilon=1+{\cal{O}}(y^{2})$, we find:
\begin{equation}\label{15}
A_{2\varphi}(\rho ,x,Q^{2})=A(x,Q^{2})+{\cal{O}}(y^{2}), \qquad \qquad \qquad \qquad
A_{\varphi}(\rho ,x,Q^{2})=A_{I}(x,Q^{2})+{\cal{O}}(y^{2}).
\end{equation}
So, like the $\sigma _{2}(x,Q^{2})$ cross section in the $\varphi$-independent case, it is the
parameters $A(x,Q^{2})$ and $A_{I}(x,Q^{2})$ that can effectively be measured in the
azimuth-dependent production.

In this paper we concentrate on the azimuthal asymmetry $A(x,Q^{2})$ associated with the
$\cos2\varphi$-distribution. We have calculated the IC contribution to the asymmetry which is
described at the parton level by the photon-quark scattering (QS) mechanism given in
Fig.~\ref{Fg.2}. Our main result can be formulated as follows:
\begin{itemize}
\item[$\star$] Contrary to the basic GF component, the IC mechanism is practically
$\cos2\varphi$-independent. This is due to the fact that the QS contribution to the
$\sigma_{A}(x,Q^{2})$ cross section is absent (for the kinematic reason) at LO and is negligibly
small (of the order of $1\%$) at NLO.
\end{itemize}
As to the $\varphi$-independent cross sections, our parton level calculations have been compared
with the previous results for the IC contribution to $\sigma_{2}(x,Q^{2})$ and
$\sigma_{L}(x,Q^{2})$ presented in Refs.~\cite{HM,KS}. Apart from two trivial misprints uncovered
in \cite{HM} for $\sigma_{L}(x,Q^{2})$, a complete agreement between all the considered results is
found.

Since the GF and QS mechanisms have strongly different $\cos2\varphi$-distributions, we investigate
the possibility to discriminate between their contributions using the azimuthal asymmetry
$A(x,Q^{2})$. We analyze separately the nonperturbative IC in the framework of the FFNS and the
perturbative IC within the VFNS.

The following properties of the nonperturbative IC contribution to the azimuthal asymmetry within
the FFNS are found:
\begin{itemize}
\item  The nonperturbative IC is practically invisible at low $x$, but affects essentially the GF
predictions at large $x$. The dominance of the $\cos2\varphi$-independent IC component at large $x$
leads to a more rapid (in comparison with the GF predictions) decreasing of $A(x,Q^{2})$ with
growth of $x$.%
\item  Contrary to the production cross sections, the $\cos 2\varphi$ asymmetry in charm  azimuthal
distributions is practically insensitive to radiative corrections at $Q^{2}\sim m^{2}$.
Perturbative stability of the combined GF+QS result for $A(x,Q^{2})$ is mainly due to the
cancellation of large NLO corrections in Eq.~(\ref{12}).%
\item  pQCD predictions for the $\cos 2\varphi$ asymmetry are parametrically stable; the GF+QS
contribution to $A(x,Q^{2})$ is practically insensitive to most of the standard uncertainties in
the QCD input parameters: $\mu _{R}$, $\mu _{F}$, $\Lambda _{QCD}$ and PDFs.%
\item  Nonperturbative corrections to the charm azimuthal asymmetry due to the gluon transverse
motion in the target are of the order of $20\%$ at $Q^{2}\leq m^{2}$ and rapidly vanish at $Q^{2}>
m^{2}$.
\end{itemize}
We conclude that the contributions of both GF and IC components to the $\cos 2\varphi$ asymmetry in
charm leptoproduction are quantitatively well defined in the FFNS: they are stable, both
parametrically and perturbatively, and insensitive (at $Q^{2}> m^{2}$) to the gluon transverse
motion in the proton. At large Bjorken $x$, the $A(x,Q^{2})$ asymmetry could be a sensitive probe
of the nonperturbative IC.

The perturbative IC has been considered within the VFNS proposed in Refs.~\cite{ACOT,collins}. The
following features of the azimuthal asymmetry should be emphasized:
\begin{itemize}
\item[$\ast$]  Contrary to the nonperturbative IC component, the perturbative one is significant
practically at all values of Bjorken $x$ and $Q^{2}>m^{2}$.%
\item[$\ast$]  The charm densities of the recent CTEQ and MRST sets of PDFs  lead to a sizeable
reduction (by about 1/3) of the GF predictions for the $\cos2\varphi$ asymmetry.
\end{itemize}
We conclude that impact of the perturbative IC on the $\cos2\varphi$ asymmetry is sizeable in the
whole region of $x$ and, for this reason, can easily be detected.

Concerning the experimental aspects, azimuthal asymmetries in charm leptoproduction can, in
principle, be measured in the COMPASS experiment at CERN, as well as in future studies at the
proposed eRHIC \cite{eRHIC,EIC} and LHeC \cite{LHeC} colliders at BNL and CERN, correspondingly.

The paper is organized as follows. In Section~\ref{part} we analyze the QS and GF parton level
predictions for the $\varphi$-dependent charm leptoproduction in the single-particle inclusive
kinematics. In particular, we discuss our results for the NLO QS cross sections and compare them
with available calculations. Hadron level predictions for $A(x,Q^{2})$ are given in
Section~\ref{hadr}. We consider the IC contributions to the asymmetry within the FFNS and VFNS in a
wide region of $x$ and $Q^{2}$. Some details of our calculations of the QS cross sections are
presented in Appendix~\ref{virt}. An overview of the soft-gluon resummation for the photon-gluon
fusion mechanism is given in Appendix~\ref{soft}. Some experimental facts in favor of the
nonperturbative IC are briefly listed in Appendix~\ref{exp}.

\section{\label{part} Partonic Cross Sections}
\subsection{Quark Scattering Mechanism}
\begin{figure}
\begin{center}
\mbox{\epsfig{file=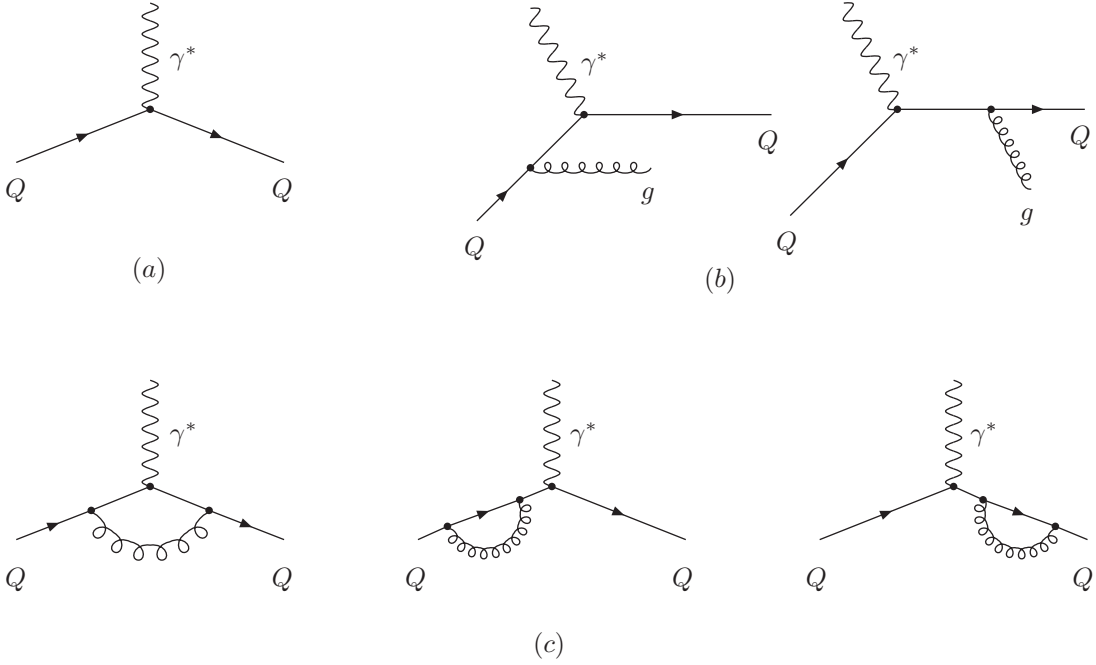,width=445pt}}
\end{center}
\caption{\label{Fg.2}\small The LO (a) and NLO (b and c) photon-quark scattering diagrams.}
\end{figure}
The momentum assignment of the deep inelastic lepton-quark scattering will be denoted as
\begin{equation}
l(\ell )+Q(k_{Q})\rightarrow l(\ell -q)+Q(p_{Q})+X(p_{X}). \label{17}
\end{equation}
Taking into account the target mass effects, the corresponding partonic cross section can be
written as follows \cite{we6}
\begin{equation}\label{18}
\frac{\text{d}^{3}\hat{\sigma}_{lQ}}{\text{d}z\text{d}Q^{2}\text{d}\varphi }=\frac{\alpha
_{em}}{(2\pi )^{2}}\frac{y^2}{z Q^{2}}\frac{\sqrt{1+4\lambda
z^{2}}}{1-\hat{\varepsilon}}\left[\hat{\sigma}_{2,Q}(z,\lambda)-
(1-\hat{\varepsilon})\hat{\sigma}_{L,Q}(z,\lambda)+
\hat{\varepsilon}\hat{\sigma}_{A,Q}(z,\lambda)\cos 2\varphi+
2\sqrt{\hat{\varepsilon}(1+\hat{\varepsilon})}\hat{\sigma}_{I,Q}(z,\lambda)\cos \varphi\right].
\end{equation}
In Eq.~(\ref{18}), we use the following definition of partonic kinematic variables:
\begin{equation}\label{19}
y=\frac{q\cdot k_{Q}}{ \ell\cdot k_{Q} },\qquad \qquad z=\frac{Q^{2}}{2q\cdot k_{Q}},\qquad
\qquad\lambda =\frac{m^{2}}{Q^{2}}.
\end{equation}
In the massive case, the (virtual) photon polarization parameter, $\hat{\varepsilon}$, has the form
\cite{we6}
\begin{equation}\label{20}
\hat{\varepsilon}=\frac{2(1-y-\lambda z^{2}y^{2})}{1+(1-y)^2+2\lambda z^{2}y^{2}}.
\end{equation}
At leading order, ${\cal O}(\alpha _{em})$, the only quark scattering subprocess is
\begin{equation}
\gamma ^{*}(q)+Q(k_{Q})\rightarrow Q(p_{Q}).  \label{21}
\end{equation}
The $\gamma ^{*}Q$ cross sections, $\hat{\sigma}_{k,Q}^{(0)}$ ($k=2,L,A,I$), corresponding to the
Born diagram (see Fig.~\ref{Fg.2}a) are:
\begin{eqnarray}
\hat{\sigma}_{2,Q}^{(0)}(z,\lambda)&=&\hat{\sigma}_{B}(z)\sqrt{1+4\lambda z^{2}}\,\delta(1-z), \nonumber\\
\hat{\sigma}_{L,Q}^{(0)}(z,\lambda)&=&\hat{\sigma}_{B}(z)\frac{4\lambda z^{2}}{\sqrt{1+4\lambda
z^{2}}}\,\delta(1-z), \label{22}\\
\hat{\sigma}_{A,Q}^{(0)}(z,\lambda)&=&\hat{\sigma}_{I,Q}^{(0)}(z,\lambda)=0,\nonumber
\end{eqnarray}
with
\begin{equation}\label{23}
\hat{\sigma}_{B}(z)=\frac{(2\pi)^2e_{Q}^{2}\alpha _{em}}{Q^{2}}\,z,
\end{equation}
where $e_{Q}$ is the quark charge in units of electromagnetic coupling constant.

To take into account the NLO ${\cal O}(\alpha _{em}\alpha _{s})$ contributions, one needs to
calculate the virtual corrections to the Born process  (given in Fig.~\ref{Fg.2}c) as well as the
real gluon emission (see Fig.~\ref{Fg.2}b):
\begin{equation}
\gamma ^{*}(q)+Q(k_{Q})\rightarrow Q(p_{Q})+g(p_{g}).  \label{24}
\end{equation}

The NLO $\varphi$-dependent cross sections, $\hat{\sigma}_{A,Q}^{(1)}$ and
$\hat{\sigma}_{I,Q}^{(1)}$, are described by the real gluon emission only. Corresponding
contributions are free of any type of singularities and the quantities $\hat{\sigma}_{A,Q}^{(1)}$
and $\hat{\sigma}_{I,Q}^{(1)}$ can be calculated directly in four dimensions.

In the $\varphi$-independent case, $\hat{\sigma}_{2,Q}^{(1)}$ and $\hat{\sigma}_{L,Q}^{(1)}$, we
also work in four dimensions. The virtual contribution (Fig.~\ref{Fg.2}c) contains ultraviolet (UV)
singularity that is removed using the on-mass-shell regularization scheme. In particular, we
calculate the absorptive part of the Feynman diagram which has no UV divergences.  The real part is
then obtained by using the appropriate dispersion relations. As to the infrared (IR) singularity,
it is regularized with the help of an infinitesimal gluon mass. This IR divergence is cancelled
when we add the bremsstrahlung contribution (Fig.~\ref{Fg.2}b). Some details of our calculations
are given in Appendix \ref{virt}.

The final (real+virtual) results for $\gamma ^{*}Q$ cross sections can be cast into the following
form:
\begin{eqnarray}
\hat{\sigma}_{2,Q}^{(1)}(z,\lambda)=\frac{\alpha_{s}}{2\pi}C_{F}\hat{\sigma}_{B}(1)
\sqrt{1+4\lambda}\,\delta(1-z)\Bigl\{-2+4\ln\lambda-\sqrt{1+4\lambda }\,\ln r+
\frac{1+2\lambda}{\sqrt{1+4\lambda}}\Bigl[2\text{Li}_{2}(r^{2})+4\text{Li}_{2}(-r)&& \nonumber\\
+3\ln^{2}(r)-4\ln r+4\ln r \ln(1+4\lambda)-2\ln r\ln\lambda\Bigr]\Bigr\}&& \nonumber \\
+\frac{\alpha_{s}}{4\pi}C_{F}\hat{\sigma}_{B}(z)\frac{1}{(1+4\lambda z^{2})^{3/2}}\biggl\{
\frac{1}{\left[1-(1-\lambda)z\right]^{2}}\Bigl[1-3z-4z^{2}+6z^{3}+8z^{4}-8z^{5} \qquad \qquad \qquad \;&& \nonumber \\
+6\lambda z\left(3-18z+13z^{2}+10z^{3}-8z^{4}\right) \qquad \qquad&&  \nonumber \\
+4\lambda^{2}z^{2}\left(8-77z+65z^{2}-2z^{3}\right)\biggr. \qquad \qquad \qquad \; \,&&  \label{25} \\
+16\lambda^{3}z^{3}\left(1-21z+12z^{2}\right)-128\lambda^{4}z^{5}\Bigr] \qquad \quad \; \,&&   \nonumber \\
+\frac{2\ln D(z,\lambda)}{\sqrt{1+4\lambda z^{2}}}\Bigl[-\left(1+z+2z^{2}+2z^{3}\right)+2\lambda
z\left(2-11z-11z^{2}\right)+8\lambda^{2}z^{2}\left(1-9z\right)\Bigr]&& \nonumber \\
-\frac{8(1+4\lambda)^{2}z^{4}}{\left(1-z\right)_{+}}-
\frac{4(1+2\lambda)(1+4\lambda)^{2}z^{4}}{\sqrt{1+4\lambda z^{2}}}\frac{\ln
D(z,\lambda)}{\left(1-z\right)_{+}}\biggr\},&& \nonumber
\end{eqnarray}
\begin{eqnarray}
\hat{\sigma}_{L,Q}^{(1)}(z,\lambda)=\frac{\alpha_{s}}{\pi}C_{F}\hat{\sigma}_{B}(1)
\frac{2\lambda}{\sqrt{1+4\lambda}}\delta(1-z)\Bigl\{-2+4\ln\lambda-\frac{4\lambda}{\sqrt{1+4\lambda
}}\,\ln r+\frac{1+2\lambda}{\sqrt{1+4\lambda}}\Bigl[2\text{Li}_{2}(r^{2})+4\text{Li}_{2}(-r)&& \nonumber\\
+3\ln^{2}(r)-4\ln r+4\ln r \ln(1+4\lambda)-2\ln r\ln\lambda\Bigr]\Bigr\}&& \nonumber \\
+\frac{\alpha_{s}}{\pi}C_{F}\hat{\sigma}_{B}(z)\frac{1}{(1+4\lambda z^{2})^{3/2}}\biggl\{
\frac{z}{\left[1-(1-\lambda)z\right]^{2}}\Bigl[(1-z)^{2}-
\lambda z\left(13-19z-2z^{2}+8z^{3}\right)\Bigr. \qquad \qquad&&   \nonumber \\
-2\lambda^{2}z^{2}\left(31-39z+8z^{2}\right)\Bigr. \qquad \qquad \qquad \qquad  \quad \; \,&& \label{26} \\
-8\lambda^{3}z^{3}\left(10-7z\right)-32\lambda^{4}z^{4}\Bigr]\Bigr.  \qquad \qquad \qquad  \qquad&&   \nonumber \\
-\frac{2\lambda z^{2}\ln D(z,\lambda)}{\sqrt{1+4\lambda z^{2}}}\left[3+3z+16\lambda z\right]\Bigr.
\qquad \qquad \qquad \qquad \; \,&& \nonumber \\
-\frac{8\lambda(1+4\lambda)z^{4}}{\left(1-z\right)_{+}}-
\frac{4\lambda(1+2\lambda)(1+4\lambda)z^{4}}{\sqrt{1+4\lambda z^{2}}}\frac{\ln
D(z,\lambda)}{\left(1-z\right)_{+}}\biggr\},&& \nonumber
\end{eqnarray}
\begin{equation}\label{27}
\hat{\sigma}_{A,Q}^{(1)}(z,\lambda)=\frac{\alpha_{s}}{2\pi}C_{F}\hat{\sigma}_{B}(z)\frac{z(1-z)}{(1+4\lambda
z^{2})^{3/2}}\biggl\{ \frac{1}{\left[1-(1-\lambda)z\right]}\left[1+2\lambda(4-3z)+8\lambda^2
z\right]+\frac{2\lambda \ln D(z,\lambda)}{\sqrt{1+4\lambda z^{2}}}\left[2+z+4\lambda
z\right]\biggr\},
\end{equation}
\begin{eqnarray}
\hat{\sigma}_{I,Q}^{(1)}(z,\lambda)=\frac{\alpha_{s}}{8\sqrt{2}}C_{F}\hat{\sigma}_{B}(z)
\frac{1}{(1+4\lambda z^{2})^{2}}\frac{\sqrt{z}}{\left[1-(1-\lambda)z\right]^{3/2}}\biggl\{
-(1-z)(1+2z)-4\lambda z\left(10-10z-z^{2}+2z^{3}\right)  \qquad \quad \; \;&& \nonumber \\
-8\lambda^{2}z^{2}\left(25-29z+8z^{2}\right)-96\lambda^{3}z^{3}\left(3-2z\right)-
128\lambda^{4}z^{4}\biggr.\;&& \label{28} \\
+8\sqrt{\lambda z\left[1-(1-\lambda)z\right]}\left[1-z^{2}+\lambda
z(13-11z)+4\lambda^{2}z^{2}(7-4z)+16\lambda^{3}z^{3}\right]\biggr\}.&& \nonumber
\end{eqnarray}
In Eqs.~(\ref{25}-\ref{28}), $C_{F}=(N_{c}^{2}-1)/(2N_{c})$, where $N_{c}$ is number of colors,
while
\begin{equation}\label{29}
D(z,\lambda)=\frac{1+2\lambda z -\sqrt{1+4\lambda z^{2}}}{1+2\lambda z +\sqrt{1+4\lambda
z^{2}}},\qquad \qquad \qquad \qquad \qquad
r=\sqrt{D(z=1,\lambda)}=\frac{\sqrt{1+4\lambda}-1}{\sqrt{1+4\lambda}+1}.
\end{equation}
The so-called "plus" distributions are defined by
\begin{equation}\label{30}
\left[g(z)\right]_{+}=g(z)-\delta(1-z)\int\limits_{0}^{1}\text{d}\zeta\,g(\zeta).
\end{equation}
For any sufficiently regular test function $h(z)$, Eq.~(\ref{30}) gives
\begin{equation}\label{31}
\int\limits_{a}^{1}\text{d}z\,h(z)\left[\frac{\ln^{k}(1-z)}{1-z}\right]_{+}=
\int\limits_{a}^{1}\text{d}z\frac{\ln^{k}(1-z)}{1-z}\left[h(z)-h(1)\right]+
h(1)\frac{\ln^{k+1}(1-a)}{k+1}.
\end{equation}

To perform a numerical investigation of the inclusive partonic cross sections, $\hat{\sigma}_{k,Q}$
($k=T,L,A,I$),{\large \ } it is convenient to introduce the dimensionless coefficient functions
$c_{k,Q}^{(n,l)}$,
\begin{equation}\label{32}
\hat{\sigma}_{k,Q}(\eta ,\lambda ,\mu ^{2})=\frac{e_{Q}^{2}\alpha _{em}\alpha _{s}(\mu
^{2})}{m^{2}}\sum_{n=0}^{\infty }\left( 4\pi \alpha _{s}(\mu ^{2})\right)
^{n}\sum_{l=0}^{n}c_{k,Q}^{(n,l)}(\eta ,\lambda )\ln ^{l}\left( \frac{\mu ^{2}}{m^{2}}\right) ,
\end{equation}
where $\mu$ is a factorization scale (we use $\mu=\mu_{F}=\mu_{R}$) and the variable $\eta$
measures the distance to the partonic threshold:
\begin{equation}\label{33}
\eta =\frac{s}{m^{2}}-1=\frac{1-z}{\lambda z},\qquad \qquad \qquad \qquad \qquad s =(q+k_{Q})^{2}.
\end{equation}

\begin{figure}
\begin{center}
\begin{tabular}{cc}
\mbox{\epsfig{file=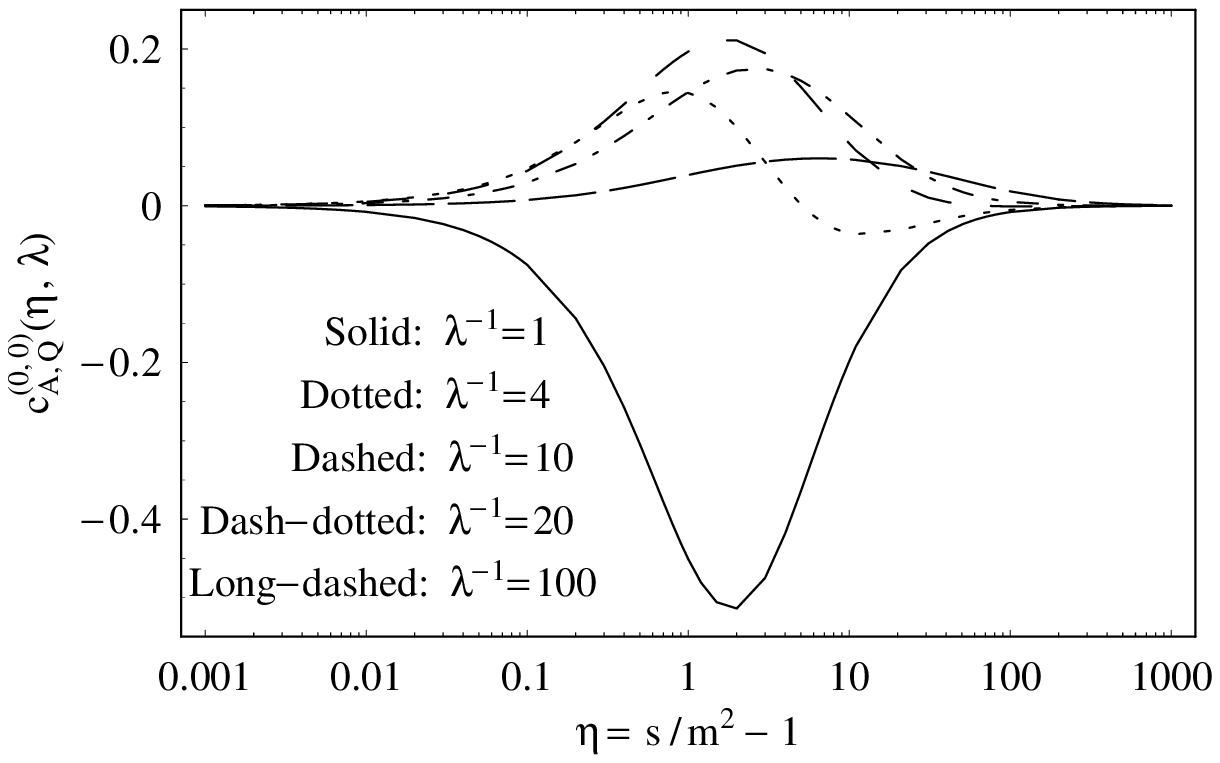,width=250pt}}
& \mbox{\epsfig{file=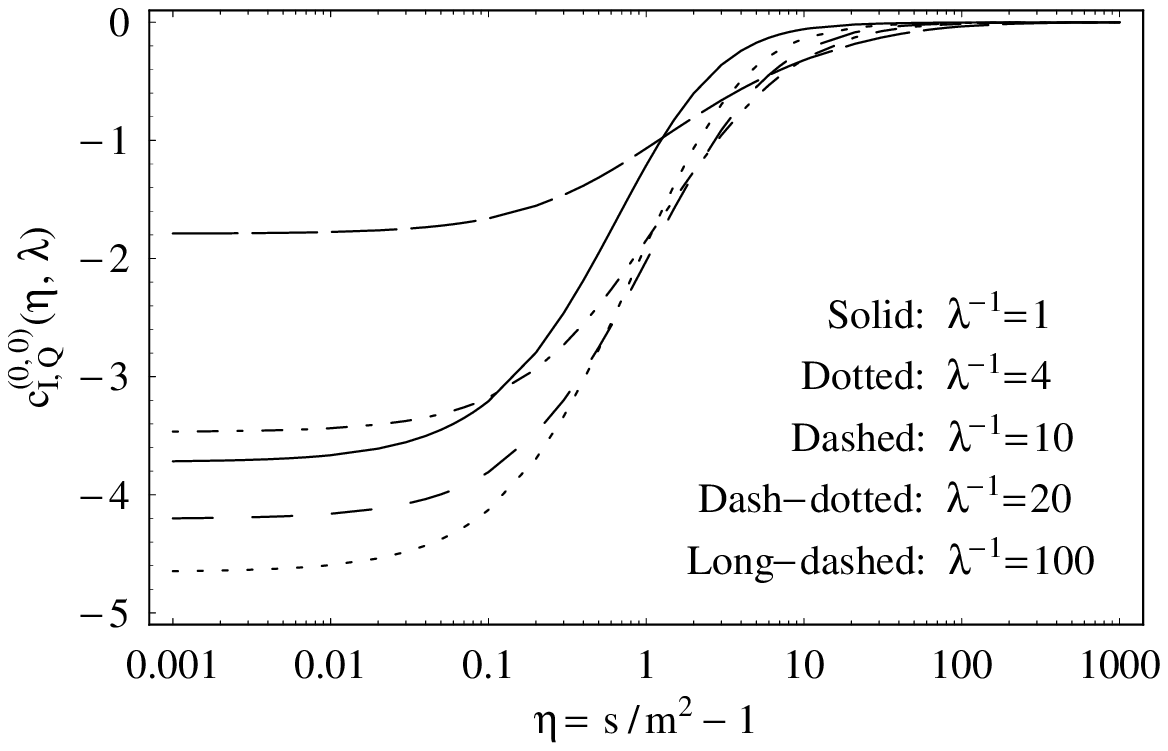,width=253pt}}\\
\end{tabular}
\caption{\label{Fg.3}\small  $c_{A,Q}^{(0,0)}(\eta,\lambda )$ and $c_{I,Q}^{(0,0)}(\eta,\lambda )$
coefficient functions at several values of $\lambda$.}
\end{center}
\end{figure}

Our analysis of the quantity $c_{A,Q}^{(0,0)}(\eta ,\lambda)$ is given in Fig.~\ref{Fg.3}. One can
see that $c_{A,Q}^{(0,0)}$ is negative at low $Q^{2}$ ($\lambda^{-1}\lesssim 1$) and positive at
high $Q^{2}$ ($\lambda^{-1}> 20$). For the intermediate values of $Q^{2}$, $c_{A,Q}^{(0,0)}(\eta
,\lambda)$ is an alternating function of $\eta$.

Our results for the coefficient function $c_{I,Q}^{(0,0)}(\eta,\lambda )$ at several values of
$\lambda$ are presented in Fig.~\ref{Fg.3}. It is seen that $c_{I,Q}^{(0,0)}$ is negative at all
values of $\eta$ and $\lambda$. Note also the threshold behavior of the coefficient function:
\begin{equation}\label{34}
c_{I,Q}^{(0,0)}(\eta\rightarrow 0,\lambda )=-\sqrt{2}\pi^{2}C_{F}\frac{\sqrt{\lambda}}{1+4\lambda}+
{\cal{O}}(\eta).
\end{equation}
This quantity takes its minimum value at $\lambda_{m}=1/4$: $c_{I,Q}^{(0,0)}(\eta = 0,\lambda_{m})
=-\pi^{2}C_{F}/\left(2\sqrt{2}\right)$.

\subsection{Comparison with Available Results}
For the first time, the NLO ${\cal O}(\alpha_{em}\alpha_{s})$ corrections to the
$\varphi$-independent IC contribution have been calculated a long time ago by Hoffmann and Moore
(HM) \cite{HM}. However, authors of Ref.~\cite{HM} don't give explicitly their definition of the
partonic cross sections that leads to a confusion in interpretation of the original HM results. To
clarify the situation, we need first to derive the relation between the lepton-quark DIS cross
section, $\text{d}\hat{\sigma}_{lQ}$, and the partonic cross sections, $\sigma^{(2)}$ and
$\sigma^{(L)}$, used in \cite{HM}. Using Eqs.~(C.1) and (C.5) in Ref.~\cite{HM}, one can express
the HM tensor $\sigma_{R}^{\mu\nu}$ in terms of "our" cross sections $\hat{\sigma}_{2,Q}$ and
$\hat{\sigma}_{L,Q}$ defined by Eq.~(\ref{18}) in the present paper. Comparing the obtained results
with the corresponding definition of $\sigma_{R}^{\mu\nu}$ via the HM cross sections $\sigma^{(2)}$
and $\sigma^{(L)}$ (given by Eqs.~(C.16) and (C.17) in Ref.~\cite{HM}), we find that
\begin{eqnarray}
\hat{\sigma}_{2,Q}(z,\lambda)&\equiv &\hat{\sigma}_{B}(z)\sqrt{1+4\lambda
z^{2}}\,\sigma^{(2)}(z,\lambda),  \label{35} \\
\hat{\sigma}_{L,Q}(z,\lambda)&\equiv &\frac{2\hat{\sigma}_{B}(z)}{\sqrt{1+4\lambda
z^{2}}}\left[\sigma^{(L)}(z,\lambda)+2\lambda z^{2}\sigma^{(2)}(z,\lambda)\right]. \label{36}
\end{eqnarray}
Now we are able to compare our results with original HM ones. It is easy to see that the LO cross
sections (defined by Eqs.~(37) in \cite{HM} and Eqs.~(\ref{22}) in our paper) obey both above
identities.  Comparing with each other the quantities $\sigma^{(2)}_{1}$ and
$\hat{\sigma}_{2,Q}^{(1)}$ (given by Eq.~(51) in \cite{HM} and Eq.~(\ref{25}) in this paper,
respectively), we find that identity (\ref{35}) is satisfied at NLO too. The situation with
longitudinal cross sections is more complicated. We have uncovered two misprints in the NLO
expression for $\sigma^{(L)}$ given by Eq.~(52) in \cite{HM}. First, the r.h.s. of this Eq. must be
multiplied by $z$. Second, the sign in front of the last term (proportional to $\delta (1-z)$) in
Eq.~(52) in Ref.~\cite{HM} must be changed \footnote{Note that this term originates from virtual
corrections and the virtual part of the longitudinal cross section given by Eq.~(39) in
Ref.~\cite{HM} also has wrong sign. See Appendix \ref{virt} for more details.}. Taking into account
these typos, we find that relation (\ref{36}) holds at NLO as well. So, our calculations of
$\hat{\sigma}_{2,Q}$ and $\hat{\sigma}_{L,Q}$ agree with the HM results.

Recently, the heavy quark initiated contributions to the $\varphi$-independent DIS structure
functions, $F_{2}$ and $F_{L}$, have been calculated by Kretzer and Schienbein (KS) \cite{KS}. The
final KS results are expressed in terms of the parton level structure functions $\hat{H}^{q}_{1}$
and $\hat{H}^{q}_{2}$. Using the definition of $\hat{H}^{q}_{1}$ and $\hat{H}^{q}_{2}$ given by
Eqs.~(7, 8) in Ref.~\cite{KS}, we obtain that
\begin{equation}\label{37}
\hat{\sigma}_{T,Q}(z,\lambda)\equiv
\frac{\alpha_{s}}{2\pi}\frac{\hat{\sigma}_{B}(z)}{\sqrt{1+4\lambda}}\frac{\hat{H}^{q}_{1}(\xi^{\prime},
\lambda)}{\sqrt{1+4\lambda z^{2}}},\qquad \qquad \qquad \qquad \hat{\sigma}_{2,Q}(z,\lambda)\equiv
\frac{\alpha_{s}}{2\pi}\hat{\sigma}_{B}(z)\sqrt{\frac{1+4\lambda}{1+4\lambda
z^{2}}}\,\hat{H}^{q}_{2}(\xi^{\prime},\lambda),
\end{equation}
where $\hat{\sigma}_{T,Q}=\hat{\sigma}_{2,Q}-\hat{\sigma}_{L,Q}$ and $\hat{\sigma}_{L,Q}$ are
defined by Eq.~(\ref{18}) in our paper and
$\xi^{\prime}=z\left(1+\sqrt{1+4\lambda}\right)\left/\left(1+\sqrt{1+4\lambda
z^{2}}\right)\right.$. To test identities (\ref{37}), one needs only to rewrite the NLO expressions
for the functions $\hat{H}^{q}_{1}(\xi^{\prime},\lambda)$ and
$\hat{H}^{q}_{2}(\xi^{\prime},\lambda)$ (given in Appendix C in Ref.~\cite{KS}) in terms of
variables $z$ and $\lambda$. Our analysis shows that relations (\ref{37}) hold at both LO and NLO.
Hence we coincide with the KS predictions for the $\gamma^{*}Q$ cross sections.

However, we disagree with the conclusion of Ref.~\cite{KS} that there are errors in the NLO
expression for $\sigma^{(2)}$ given in Ref.~\cite{HM} \footnote{In detail, the KS point of view on
the HM results is presented in PhD thesis \cite{KS-thesis}, pp.~158-160.}. As explained above, a
correct interpretation of the quantities $\sigma^{(2)}$ and $\sigma^{(L)}$ used in \cite{HM} leads
to a complete agreement between the HM, KS and our results for $\varphi$-independent cross
sections.

As to the $\varphi$-dependent DIS, pQCD predictions for the $\gamma^{*}Q$ cross sections
$\hat{\sigma}_{A,Q}(z,\lambda)$ and $\hat{\sigma}_{I,Q}(z,\lambda)$ in the case of arbitrary values
of $m^{2}$ and $Q^{2}$ are not, to our knowledge, available in the literature. For this reason, we
have performed several cross checks of our results against well known calculations in two limits:
$m^{2}\rightarrow 0$ and $Q^{2}\rightarrow 0$. In particular, in the chiral limit, we reproduce the
original results of Georgi and Politzer \cite{GP} and M\'{e}ndez \cite{Mendez} for
$\hat{\sigma}_{I,Q}(z,\lambda\rightarrow 0)$ and $\hat{\sigma}_{A,Q}(z,\lambda\rightarrow 0)$. In
the case of $Q^{2}\rightarrow 0$, our predictions for $\hat{\sigma}_{2,Q}(s,Q^{2}\rightarrow 0)$
and $\hat{\sigma}_{A,Q}(s,Q^{2}\rightarrow 0)$ given by Eqs.~(\ref{25},\ref{27}) reduce to the QED
textbook results for the Compton scattering of polarized photons \cite{Fano}.

\subsection{Photon-Gluon Fusion}
The gluon fusion component of the semi-inclusive DIS is the following parton level interaction:
\begin{equation}
l(\ell )+g(k_{g})\rightarrow l(\ell -q)+Q(p_{Q})+X[\overline{Q}](p_{X}). \label{d1}
\end{equation}
Corresponding lepton-gluon cross section, $\text{d}\hat{\sigma}_{lg}$, has the following
decomposition in terms of the helicity $\gamma ^{*}g$ cross sections:
\begin{equation}\label{d2}
\frac{\text{d}^{3}\hat{\sigma}_{lg}}{\text{d}z\text{d}Q^{2}\text{d}\varphi }=\frac{\alpha
_{em}}{(2\pi )^{2}}\frac{1}{z Q^{2}}\frac{y^2}{1-\varepsilon}\left[\hat{\sigma}_{2,g}(z,\lambda)-
(1-\varepsilon)\hat{\sigma}_{L,g}(z,\lambda)+ \varepsilon\hat{\sigma}_{A,g}(z,\lambda)\cos
2\varphi+ 2\sqrt{\varepsilon(1+\varepsilon)}\hat{\sigma}_{I,g}(z,\lambda)\cos \varphi\right],
\end{equation}
where the quantity $\varepsilon$ is defined by Eq.~(\ref{3}) with $y=\left.(q\cdot k_{g})\right
/(\ell\cdot k_{g})$.
\begin{figure}
\begin{center}
\mbox{\epsfig{file=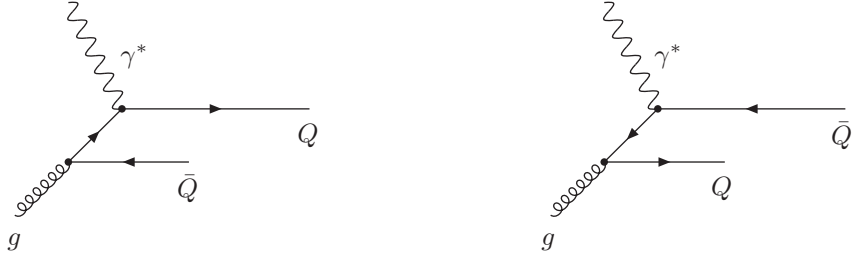,width=355pt}}
\end{center}
\caption{\label{Fg.4}\small The LO photon-gluon fusion diagrams.}
\end{figure}

At LO, ${\cal O}(\alpha_{em}\alpha_{s})$, the only gluon fusion subprocess responsible for heavy
flavor production is
\begin{equation}\label{38}
\gamma ^{*}(q)+g(k_{g})\rightarrow Q(p_{Q})+\overline{Q}(p_{\stackrel{\_}{Q}}).
\end{equation}
The $\gamma ^{*}g$ cross sections, $\hat{\sigma}_{k,g}^{(0)}$ ($k=2,L,A,I$), corresponding to the
Born diagrams given in Fig.~\ref{Fg.4} have the form \cite{LW1,Watson}:
\begin{eqnarray}
\hat{\sigma}_{2,g}^{(0)}(z,\lambda)&=&\frac{\alpha_{s}}{2\pi}\hat{\sigma}_{B}(z)
\Bigl\{\left[(1-z)^{2}+z^{2}+4\lambda z(1-3z)-8\lambda^{2}z^{2}\right]
\ln\frac{1+\beta_{z}}{1-\beta_{z}}-
\left[1+4z(1-z)(\lambda-2)\right]\beta_{z}\Bigr\},  \nonumber\\
\hat{\sigma}_{L,g}^{(0)}(z,\lambda)&=&\frac{2\alpha_{s}}{\pi}\hat{\sigma}_{B}(z)z
\Bigl\{-2\lambda z\ln\frac{1+\beta_{z}}{1-\beta_{z}}+\left(1-z\right)\beta_{z}\Bigr\},  \nonumber\\
\hat{\sigma}_{A,g}^{(0)}(z,\lambda)&=&\frac{\alpha_{s}}{\pi}\hat{\sigma}_{B}(z)z
\Bigl\{2\lambda\left[1-2z(1+\lambda)\right]\ln\frac{1+\beta_{z}}{1-\beta_{z}}+
(1-2\lambda)(1-z)\beta_{z}\Bigr\},  \label{39} \\
\hat{\sigma}_{I,g}^{(0)}(z,\lambda)&=&0,  \nonumber
\end{eqnarray}
where $\hat{\sigma}_{B}(z)$ is defined by Eq.~(\ref{23}) and the following notations are used:
\begin{equation}\label{40}
z=\frac{Q^{2}}{2q\cdot k_{g}},\qquad \qquad\lambda =\frac{m^{2}}{Q^{2}}, \qquad \qquad
\beta_{z}=\sqrt{1-\frac{4\lambda z}{1-z}}.
\end{equation}
Note that the $\cos \varphi $ dependence vanishes in the GF mechanism due to the $Q\leftrightarrow
\overline{Q}$ symmetry which, at leading order, requires invariance under $\varphi \rightarrow
\varphi +\pi$ \cite{LW2}.

As to the NLO results, presently, only $\varphi$-independent quantities $\hat{\sigma}_{2,g}^{(1)}$
and $\hat{\sigma}_{L,g}^{(1)}$ are known exactly \cite{LRSN}. For this reason, we will use in our
analysis the so-called soft-gluon approximation for the NLO $\gamma ^{*}g$ cross sections (see
Appendix \ref{soft}). As shown in Refs.~\cite{Laenen-Moch,we2,we4}, at energies not so far from the
production threshold, the soft-gluon radiation is the dominant perturbative mechanism in the
$\gamma ^{*}g$ interactions.

\section{\label{hadr}Hadron Level Results}
\subsection{\label{ha}Fixed Flavor Number Scheme and Nonperturbative Intrinsic Charm}
In the fixed flavor number scheme \footnote{This approach is sometimes referred to as the
fixed-order perturbation theory (FOPT).}, the wave function of the proton consists of light quarks
$u,d,s$ and gluons $g$. Heavy flavor production in DIS is dominated by the gluon fusion mechanism.
Corresponding hadron level cross sections, $\sigma_{k,GF}(x,\lambda)$, have the form
\begin{eqnarray}
\sigma_{k,GF}(x,\lambda)&=&\int\limits_{\chi}^{1}\text{d}z\,g(z,\mu_{F})\,\hat{\sigma}_{k,g}
\!\left(x/z,\lambda,\mu_{F}\right), \qquad \qquad \qquad (k=2,L,A,I), \label{41}\\
\chi&=&x\left(1+4\lambda\right), \label{42}
\end{eqnarray}
where $g(z,\mu _{F})$ describes gluon density in the proton evaluated at a factorization scale $\mu
_{F}$. The lowest order GF cross sections, $\hat{\sigma}_{k,g}^{(0)}$ ($k=2,L,A,I$), are given by
Eqs.~(\ref{39}). The NLO results, $\hat{\sigma}_{k,g}^{(1)}$, to the next-to-leading logarithmic
accuracy are presented in Appendix \ref{soft}.

We neglect the $\gamma ^{*}q(\bar{q})$ fusion subprocesses. This is justified as their
contributions to heavy quark leptoproduction vanish at LO and are small at NLO \cite{LRSN}.

In the FFNS, the intrinsic heavy flavor component of the proton wave function is generated by
$gg\rightarrow Q\bar{Q}$ fluctuations where the gluons are coupled to different valence quarks. In
the present paper, this component is referred to as the nonperturbative intrinsic charm (bottom).
The probability of the corresponding five-quark Fock state, $\left|uudQ\bar{Q}\right\rangle$, is of
higher twist since it scales as $\Lambda_{QCD}^{2}\left/m^{2}\right.$ \cite{polyakov}. However,
since all of the quarks tend to travel coherently at same rapidity in the
$\left|uudQ\bar{Q}\right\rangle$ bound state, the heaviest constituents carry the largest momentum
fraction. For this reason, the heavy flavor distribution function has a more "hard" $z$-behavior
than the light parton densities. Since all of the densities vanish at $z\to 1$, the hardest PDF
becomes dominant at sufficiently large $z$ independently of normalization.

Convolution of PDFs with partonic cross sections does not violate this observation. In particular,
assuming a gluon density $g(z)\sim (1-z)^n$ (where $n=3-5$), we obtain that the LO GF contribution
to $F_{2}$ scales as $(1-\chi)^{n+3/2}$ at $\chi\to 1$, where $\chi$ is defined by Eq.~(\ref{42}).
In the case of Hoffman and Moore charm density (see below), the LO IC contribution is proportional
to $(1-\chi)$ at $\chi\to 1$. It is easy to see that, independently of normalizations, the IC
contribution to be dominate over the more "soft" GF component at large enough $x$.

For the first time, the intrinsic charm momentum distribution in the five-quark state
$\left|uudc\bar{c}\right\rangle$ was derived by Brodsky, Hoyer, Peterson and Sakai (BHPS) in the
framework of a light-cone model \cite{BHPS,BPS}. Neglecting the transverse motion of constituents,
they have obtained in the heavy quark limit that
\begin{equation}\label{43}
c(z)=\frac{N_{5}}{6}z^{2}\left[6z(1+z)\ln z+(1-z)(1+10z+z^{2})\right],
\end{equation}
where $N_{5}=36$ corresponds to a $1\%$ probability for IC in the nucleon:
$\int^{1}_{0}c(z)\text{d}z=0.01$.

Hoffmann and Moore (HM) \cite{HM} incorporated mass effects in the BHPS approach. They first
introduced a mass scaling variable $\xi$,
\begin{equation}\label{44}
\xi=\frac{2ax}{1+\sqrt{1+4\lambda_{N}x^{2}}},\qquad \qquad \qquad \qquad
a=\frac{1+\sqrt{1+4\lambda}}{2},
\end{equation}
where $\lambda_{N}=m^{2}_{N}\left/Q^{2}\right.$. To provide correct threshold behavior of the charm
density, the constraint $\xi\leq\gamma<1$ was imposed where
\begin{equation}\label{45}
\gamma=\frac{2a\hat{x}}{1+\sqrt{1+4\lambda_{N}\hat{x}^{2}}},\qquad \qquad \qquad \qquad
\hat{x}=\frac{1}{1+4\lambda-\lambda_{N}}.
\end{equation}
Resulting charm distribution function, $c(\xi,\gamma)$, has the following form in the HM approach:
\begin{equation}\label{46}
c(\xi,\gamma)=\left\{%
\begin{array}{ll}
c(\xi)-\displaystyle{\frac{\xi}{\gamma}}c(\gamma),&\qquad \xi\leq\gamma\\
0,&\qquad \xi>\gamma
\end{array}%
\right.
\end{equation}
with $c(\xi)$ defined by Eq.~(\ref{43}). Corresponding hadron level cross sections for the
$(c+\bar{c})$ production, $\sigma_{k,QS}(x,\lambda)$, due to the heavy quark scattering (QS)
mechanism, are
\begin{equation}\label{47}
\sigma_{k,QS}(x,\lambda)=\int\limits_{\xi}^{\gamma}\frac{\text{d}z}{\sqrt{1+4\lambda\xi^{2}/z^{2}}}
\,c_{+}(z,\gamma)\,\hat{\sigma}_{k,c}\!\left(\xi/z,\lambda\right),\qquad \qquad \qquad (k=2,L,A,I),
\end{equation}
where the charm density $c_{+}(z,\gamma)\equiv c(z,\gamma)+\bar{c}(z,\gamma)$. The LO and NLO
expressions for the partonic cross sections $\hat{\sigma}_{k,c}(z,\lambda)$ are given by
Eqs.~(\ref{22}) and (\ref{25}-\ref{28}), respectively.

Note also that, in the FFNS, the full cross section for the charm production,
$\sigma_{k}(x,\lambda)$, is simply a sum of the GF and IC components:
\begin{equation}\label{48}
\sigma_{k}(x,\lambda)=\sigma_{k,GF}(x,\lambda)+\sigma_{k,QS}(x,\lambda), \qquad \qquad \qquad
\qquad \qquad (k=2,L,A,I).
\end{equation}
\begin{figure}
\begin{center}
\begin{tabular}{cc}
\mbox{\epsfig{file=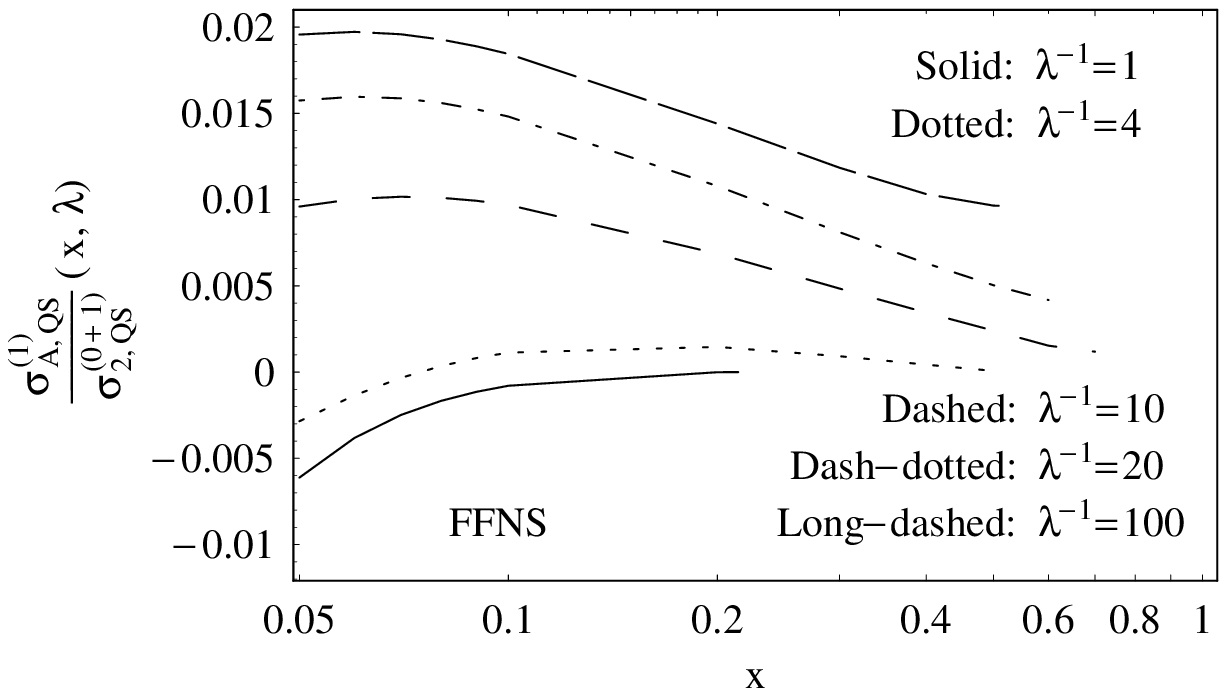,width=250pt}}
& \mbox{\epsfig{file=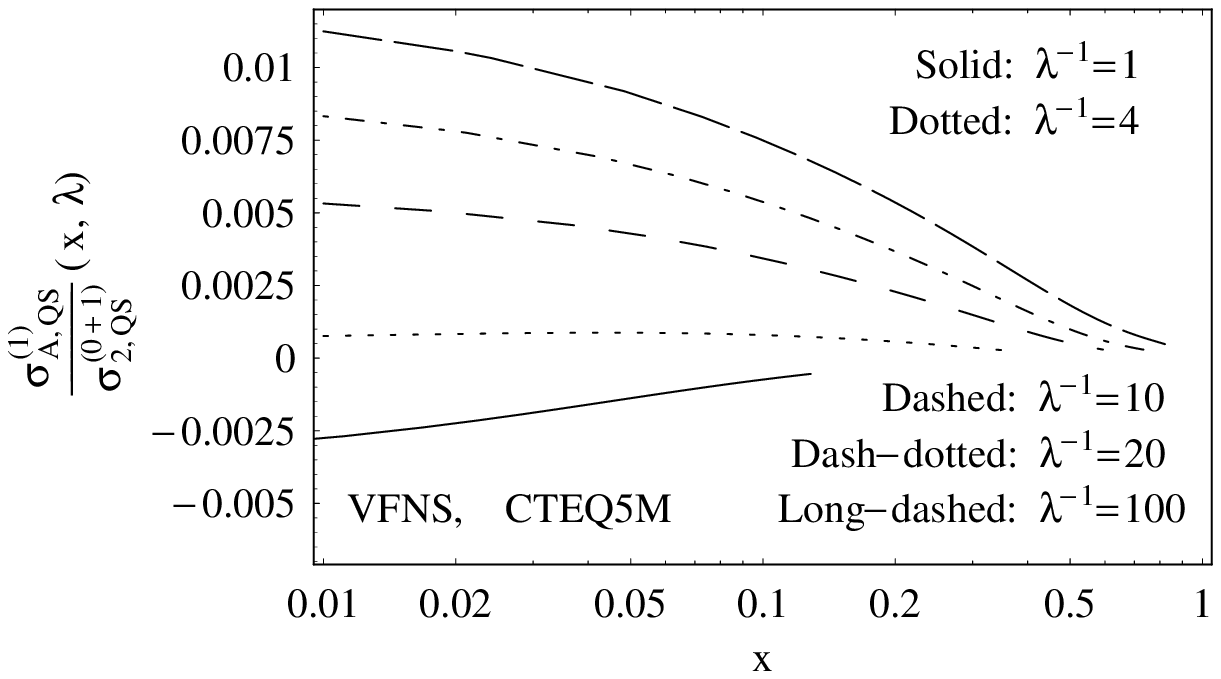,width=250pt}}\\
\end{tabular}
\caption{\label{Fg.5}\small The quantity
$\bigl(\sigma^{(1)}_{A,QS}\,\bigl/\sigma^{(0+1)}_{2,QS}\bigr)(x,\lambda)$ in the FFNS with the HM
\cite{HM} charm density (\emph{left panel}) and in the VFNS with the CTEQ5M charm distribution
function (\emph{right panel}).}
\end{center}
\end{figure}

Let us discuss the FFNS predictions for the hadron level asymmetry parameter $A(x,Q^{2})$ defined
by Eq.~(\ref{12}). First we consider the ratio
$\bigl(\sigma^{(1)}_{A,QS}\,\bigl/\sigma^{(0+1)}_{2,QS}\bigr)(x,\lambda)$, i.e., the mere IC
contribution to $A(x,Q^{2})$. In Fig.~\ref{Fg.5}, the $x$-behavior of this quantity at various
values of $\lambda$ is presented. One can see that the ratio
$\bigl(\sigma^{(1)}_{A,QS}\,\bigl/\sigma^{(0+1)}_{2,QS}\bigr)(x,\lambda)$ is negligibly small (of
the order of $1\%$) practically at all values of $Q^{2}>m^{2}$. Note that this fact is independent
of the charm density we use (see, for instance, the right panel in Fig.~\ref{Fg.5}), but is only
due to the smallness of the partonic cross section $\hat{\sigma}^{(1)}_{A,c}(z,\lambda)$
\cite{we6}. So, the quantity $\sigma_{A,QS}(x,\lambda)$ is exactly zero at LO \footnote{Remember
that the LO quantity $\hat{\sigma}^{(0)}_{A,c}(z,\lambda)$ vanishes for the kinematic reason, see
Eqs.~(\ref{22}).} and negligibly small at NLO. This implies that the IC contribution has no
practically $\cos2\varphi$-dependence and, for this reason, we will neglect both
$\hat{\sigma}_{A,c}(z,\lambda)$ and $\sigma_{A,QS}(x,\lambda)$ cross sections in our further
analysis.
\begin{figure}
\begin{center}
\begin{tabular}{cc}
\mbox{\epsfig{file=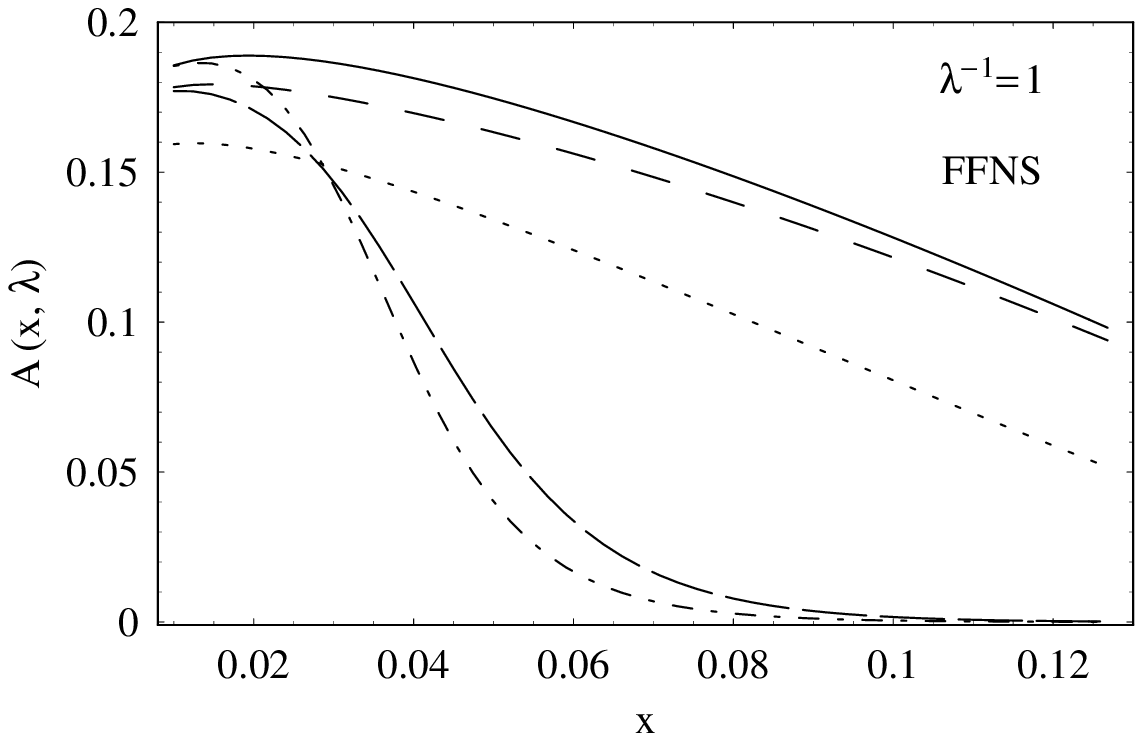,width=252pt}}
& \mbox{\epsfig{file=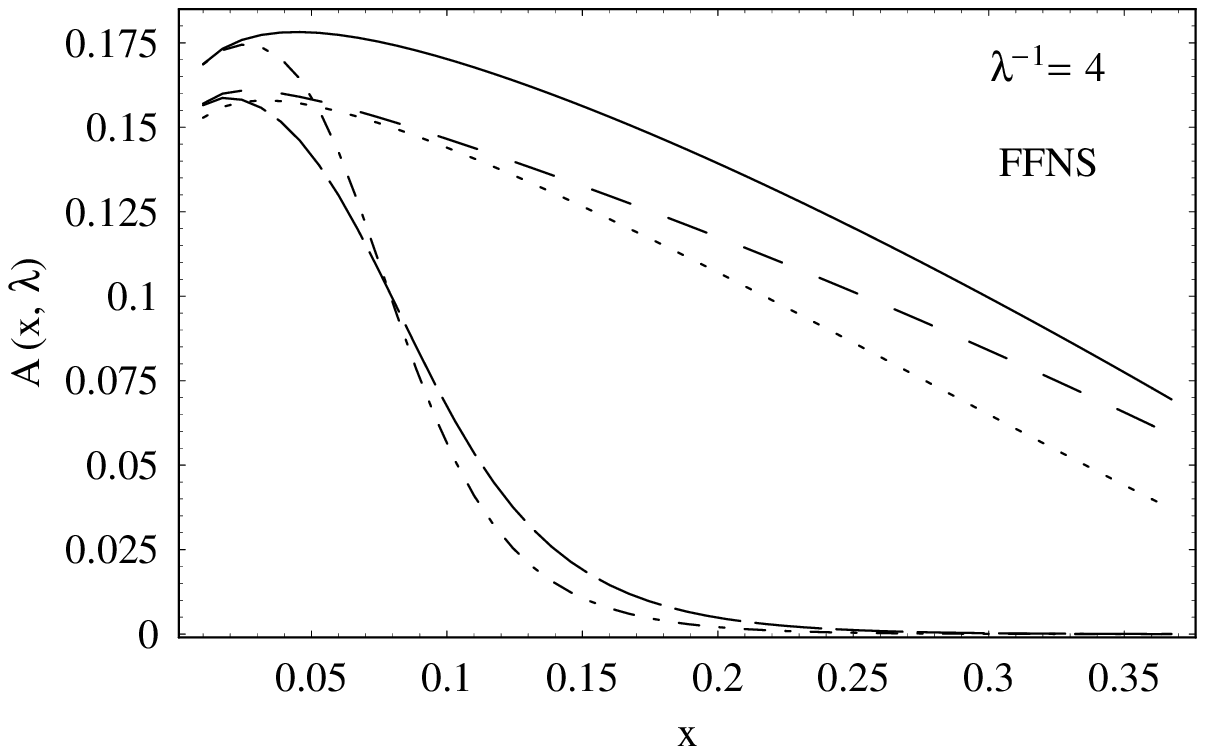,width=245pt}}\\
\mbox{\epsfig{file=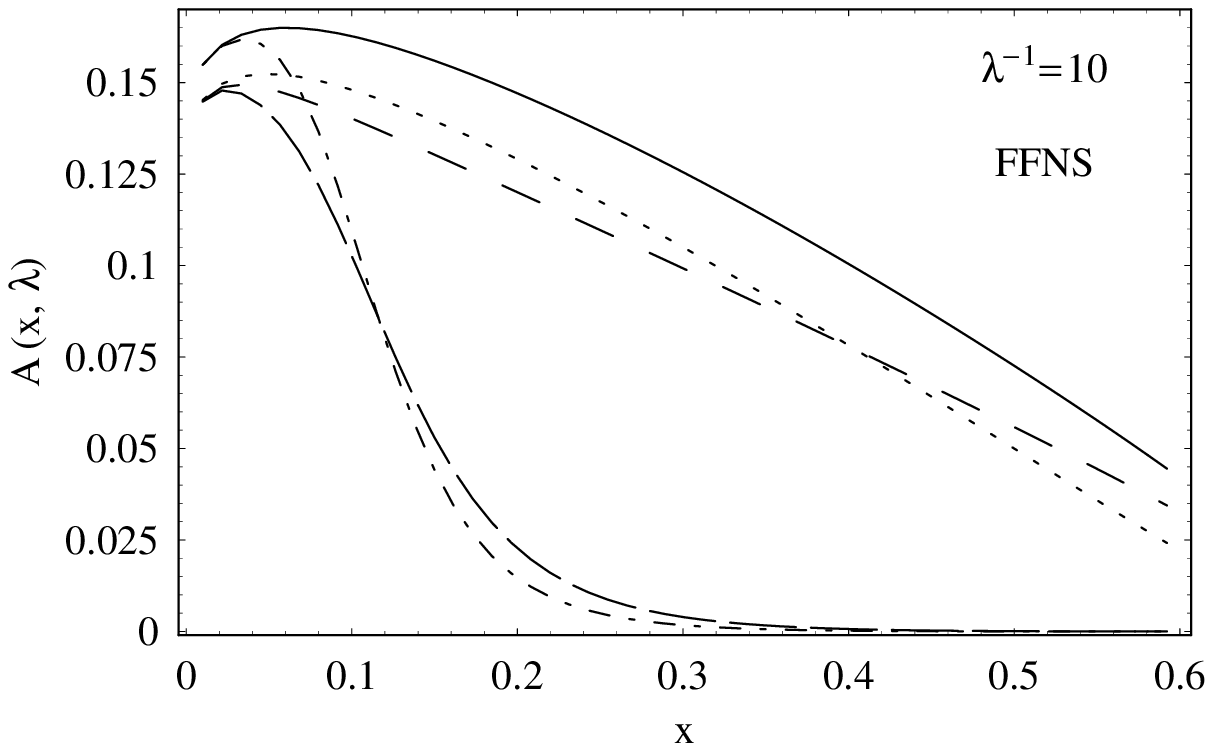,width=243pt}}
& \mbox{\epsfig{file=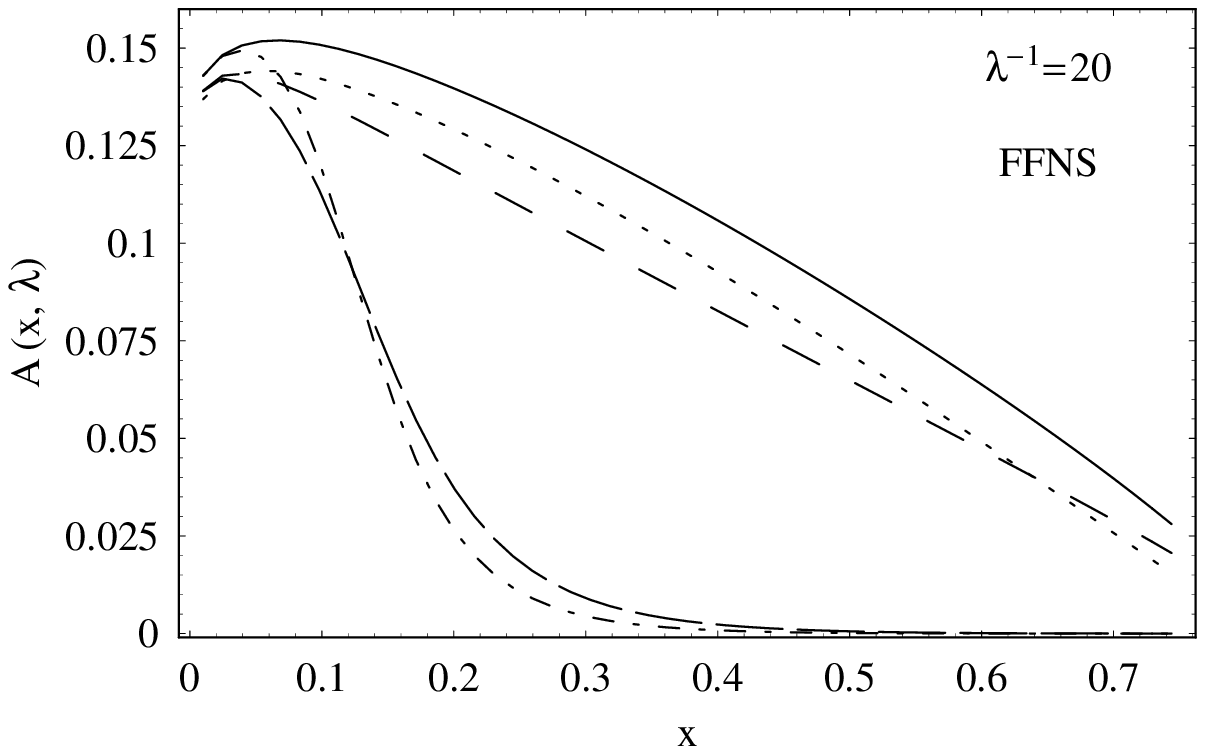,width=245pt}}\\
\mbox{\epsfig{file=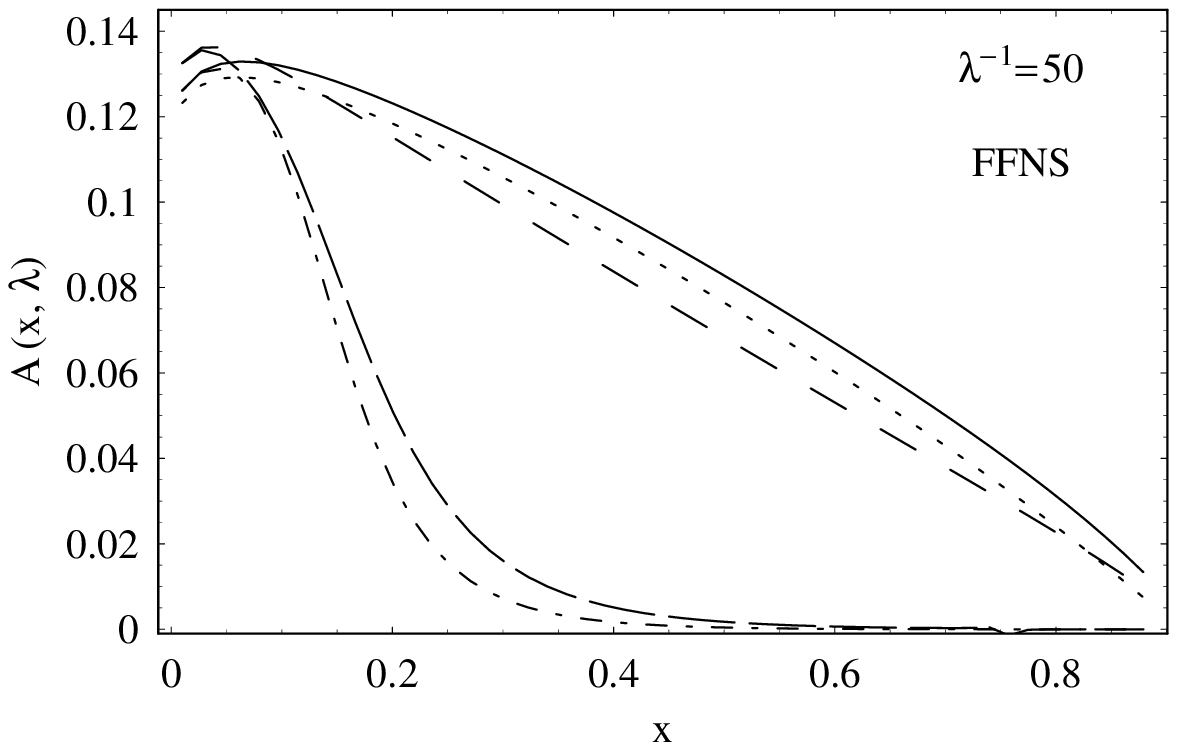,width=250pt}}
& \mbox{\epsfig{file=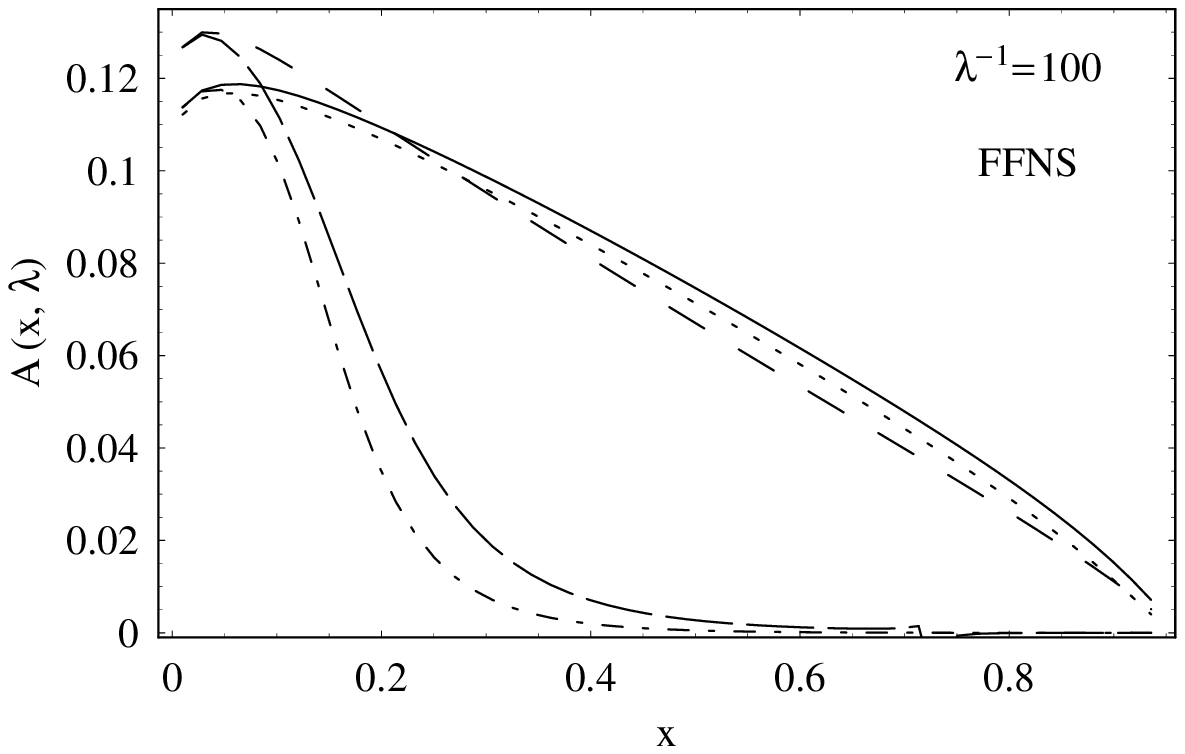,width=247pt}}\\
\end{tabular}
\caption{\label{Fg.6}\small Azimuthal asymmetry parameter $A(x,\lambda)$ in the FFNS at several
values of $\lambda$ in the case of $\int^{1}_{0}c(z)\text{d}z=1\%$. The following contributions are
plotted: $\text{GF}^{\text{(LO)}}$ (solid lines), $\text{GF}^{\text{(LO)}}$+$k_{T}$-kick (dotted
lines), $\text{GF}^{\text{(NLO)}}$ (dashed lines),
$\text{GF}^{\text{(LO)}}$+$\text{QS}^{\text{(LO)}}$ (dash-dotted lines) and
$\text{GF}^{\text{(NLO)}}$+$\text{QS}^{\text{(NLO)}}$ (long-dashed lines).}
\end{center}
\end{figure}

Fig.~\ref{Fg.6} shows $A(x,\lambda)$ as a function of $x$ for several values of variable $\lambda$:
$\lambda^{-1}=1,4,10,20,50$ and 100. We display both LO and NLO predictions  of the GF mechanism as
well as the analogous results of the combined GF+QS contribution. The azimuthal asymmetry due to
the mere LO GF component is given by solid line. The NLO GF predictions are plotted by dashed line.
The LO and NLO results of the total GF+QS contribution are given by dash-dotted and long-dashed
lines, respectively. In our calculations, the CTEQ5M \cite{CTEQ5} parametrization of the gluon
distribution function is used and a $1\%$ probability for IC in the nucleon is assumed. Throughout
this paper, the value $\mu_{F}=\mu_{R}=\sqrt{m^{2}+Q^{2}}$ for both factorization and
renormalization scales is chosen. In accordance with the CTEQ5M parametrization, we use $m_{c}=1.3
$ GeV and $\Lambda_{4}=326$ MeV \cite{CTEQ5}.

One can see from Fig.~\ref{Fg.6} the following basic features of the azimuthal asymmetry,
$A(x,\lambda)$, within the FFNS. First, as expected, the nonperturbative IC contribution is
practically invisible at low $x$, but affects essentially the GF predictions at large $x$. Since,
contrary to the GF mechanism, the QS component is practically $\cos2\varphi$-independent, the
dominance of the IC contribution at large $x$ leads to a more rapid (in comparison with the GF
predictions) decreasing of $A(x,\lambda)$ with growth of $x$.

The most remarkable property of the azimuthal asymmetry is its perturbative stability. In
Refs.~\cite{we2,we4}, the NLO soft-gluon corrections to the GF predictions for the $\cos2\varphi$
asymmetry in heavy quark photo- and leptoproduction was calculated. It was shown that, contrary to
the production cross sections, the quantity $A(x,\lambda)$ is practically insensitive to soft
radiation. One can see from Fig.~\ref{Fg.6} in the present paper that the NLO corrections to the LO
GF predictions for $A(x,\lambda)$ are about few percent at not large $x$.  This implies that large
soft-gluon corrections to $\sigma_{A,GF}^{(LO)}$ and $\sigma_{2,GF}^{(LO)}$ (increasing both cross
sections by a factor of two) cancel each other in the ratio
$\bigl(\sigma_{A,GF}^{(NLO)}\!\bigl/\sigma_{2,GF}^{(NLO)}\bigr)(x,\lambda)$ with a good accuracy.
In terms of so-called $K$-factors,
$K_{k}(x,\lambda)=\bigl(\sigma_{k}^{(NLO)}\!\bigl/\sigma_{k}^{(LO)}\bigr)(x,\lambda)$ for
$k=2,L,A,I$, perturbative stability of the GF predictions for $A(x,\lambda)$ is provided by the
fact that the corresponding $K$-factors are approximately the same at not large $x$:
$K_{A,GF}(x,\lambda)\approx K_{2,GF}(x,\lambda)$.

Comparing with each other the dash-dotted and long-dashed curves in Fig.~\ref{Fg.6}, we see that
the NLO corrections to the combined GF+QS result for $A(x,\lambda)$ are also small. In this case,
three reasons are responsible for the closeness of the LO and NLO predictions. At small $x$, where
the nonperturbative IC contribution is negligible, perturbative stability of the asymmetry is
provided by the GF component. In the large-$x$ region, where the IC mechanism dominates, the
azimuthal asymmetry rapidly vanishes with growth of $x$ at both LO and NLO because the QS component
is practically $\cos2\varphi$-independent, $\hat{\sigma}^{(1)}_{A,c}(x,\lambda)\approx\hat{\sigma
}^{(0)}_{A,c}(x,\lambda)=0$ \footnote{Although the ratio $(A^{(NLO)}/A^{(LO)})(x,\lambda)$ is
sizeable at sufficiently large $x$, the absolute values of the quantities $A^{(LO)}(x,\lambda)$ and
$A^{(NLO)}(x,\lambda)$ become so small that it seems reasonable to consider the asymmetry as
equally negligible at both LO and NLO and treat the predictions as perturbatively stable.}. At
intermediate values of $x$, where both mechanisms are essential, perturbative stability of
$A(x,\lambda)$ is due to the similarity of the corresponding $K$-factors: $K_{2,GF}(x,\lambda)\sim
K_{2,QS}(x,\lambda)$ \footnote{Note however that this similarity takes only place at intermediate
values of $x$ where both GF and QS components are essential. In the low- and large-$x$ regions, the
factors $K_{2,GF}(x,\lambda)$ and $K_{2,QS}(x,\lambda)$ are strongly different.}.

Another remarkable property of the azimuthal asymmetry closely related to fast perturbative
convergence is its parametric stability \footnote{Of course, parametric stability of the fixed
order results does not imply a fast convergence of the corresponding series. However, a fast
convergent series must be parametrically stable. In particular, it must be $\mu _{R}$- and $\mu
_{F}$-independent.}. The analysis of Refs.~\cite{we1,we4} shows that the GF predictions for the
$\cos 2\varphi $ asymmetry are less sensitive to standard uncertainties in the QCD input parameters
($m,\mu _{R},\mu _{F},\Lambda_{QCD}$ and PDFs) than the corresponding ones for the production cross
sections. We have verified that the same situation takes also place for the combined GF+QS results.

Let us discuss how the GF predictions for the azimuthal asymmetry are affected by nonperturbative
contributions due to the intrinsic transverse motion of the gluon in the target. Because of the
relatively low $c$-quark mass, these contributions are especially important in the description of
the cross sections for charmed particle production.

To introduce $k_{T}$ degrees of freedom, $\vec{k}_{g}\simeq \zeta\vec{p}+\vec{k}_{T}$, one extends
the integral over the parton distribution function in Eq. (\ref{41}) to $k_{T}$-space,
\begin{equation}  \label{49}
\text{d}\zeta\,g(\zeta,\mu _{F})\rightarrow \text{d}\zeta\,\text{d}^{2}k_{T}f\big(\vec{k}_{T}\big)
g(\zeta,\mu _{F}).
\end{equation}
The transverse momentum distribution, $f\big( \vec{k}_{T}\big) $, is usually taken to be a
Gaussian:
\begin{equation}  \label{50}
f\big( \vec{k}_{T}\big) =\frac{{\rm {e}}^{-\vec{k}_{T}^{2}/\langle k_{T}^{2}\rangle }}{\pi \langle
k_{T}^{2}\rangle }.
\end{equation}
In practice, an analytic treatment of $k_{T}$ effects is usually used. According to Ref.~\cite{kT},
the $k_{T}$-smeared differential cross section of the process (\ref{1}) is a two-dimensional
convolution:
\begin{equation}  \label{51}
\frac{\text{d}^{4}\sigma _{lN}^{{\rm {kick}}}}{\text{d}x\text{d}Q^{2}\text{d} p_{QT}\text{d}\varphi
}\left( \vec{p}_{QT}\right) =\int \text{d}^{2}k_{T} \frac{{\rm {e}}^{-\vec{k}_{T}^{2}/\langle
k_{T}^{2}\rangle }}{\pi \langle k_{T}^{2}\rangle }\frac{\text{d}^{4}\sigma
_{lN}}{\text{d}x\text{d}Q^{2} \text{d}p_{QT}\text{d}\varphi }\Big( \vec{p}_{QT}-\frac{1}{2}\vec{k}
_{T}\Big) .
\end{equation}
The factor $\frac{1}{2}$ in front of $\vec{k}_{T}$ in the r.h.s. of Eq.~(\ref {51}) reflects the
fact that the heavy quark carries away about one half of the initial energy in the reaction
(\ref{1}).

Values of the $k_{T}$-kick corrections to the LO GF predictions for the $\cos 2\varphi $ asymmetry
in the charm production are shown in Fig.~\ref{Fg.6} by dotted curves. Calculating the $k_{T}$-kick
effect we use $\langle k_{T}^{2}\rangle =0.5$ GeV$^{\text{2}}$. One can see that $k_{T}$-smearing
for $A(x,Q^{2})$ is about $20$-$25\%$ in the region of low $Q^{2}\lesssim m^{2}$ and rapidly
decreases at high $Q^{2}$.

\begin{figure}
\begin{center}
\begin{tabular}{cc}
\mbox{\epsfig{file=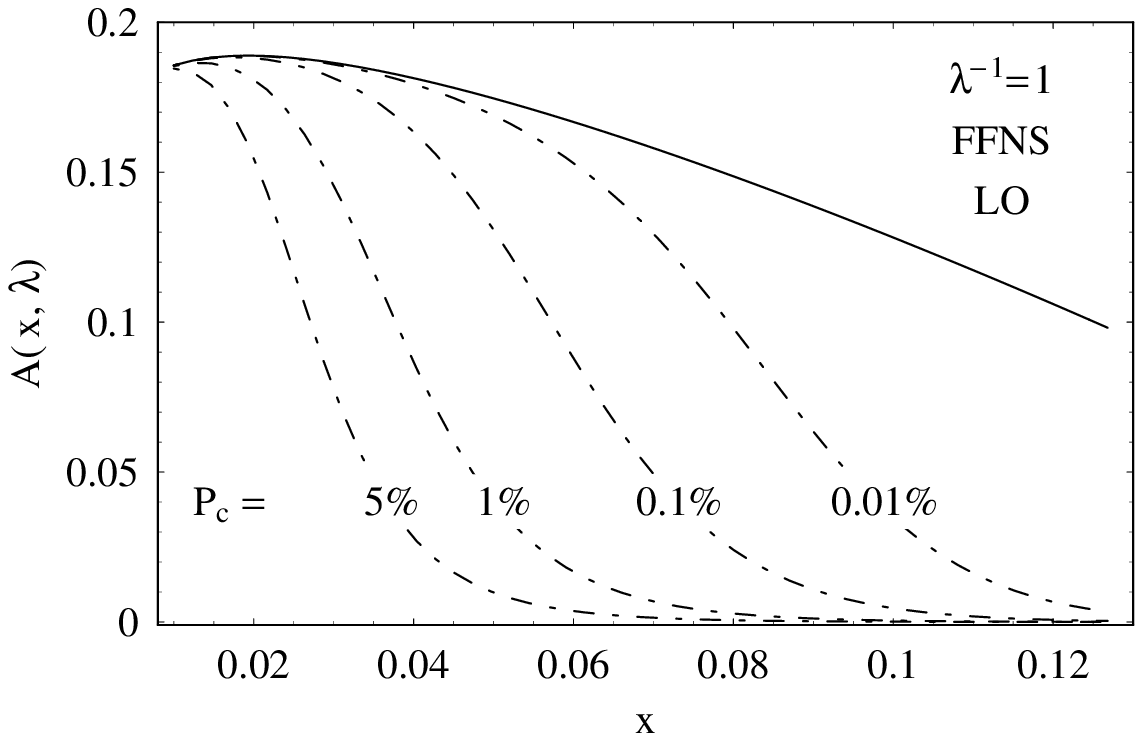,width=252pt}}
& \mbox{\epsfig{file=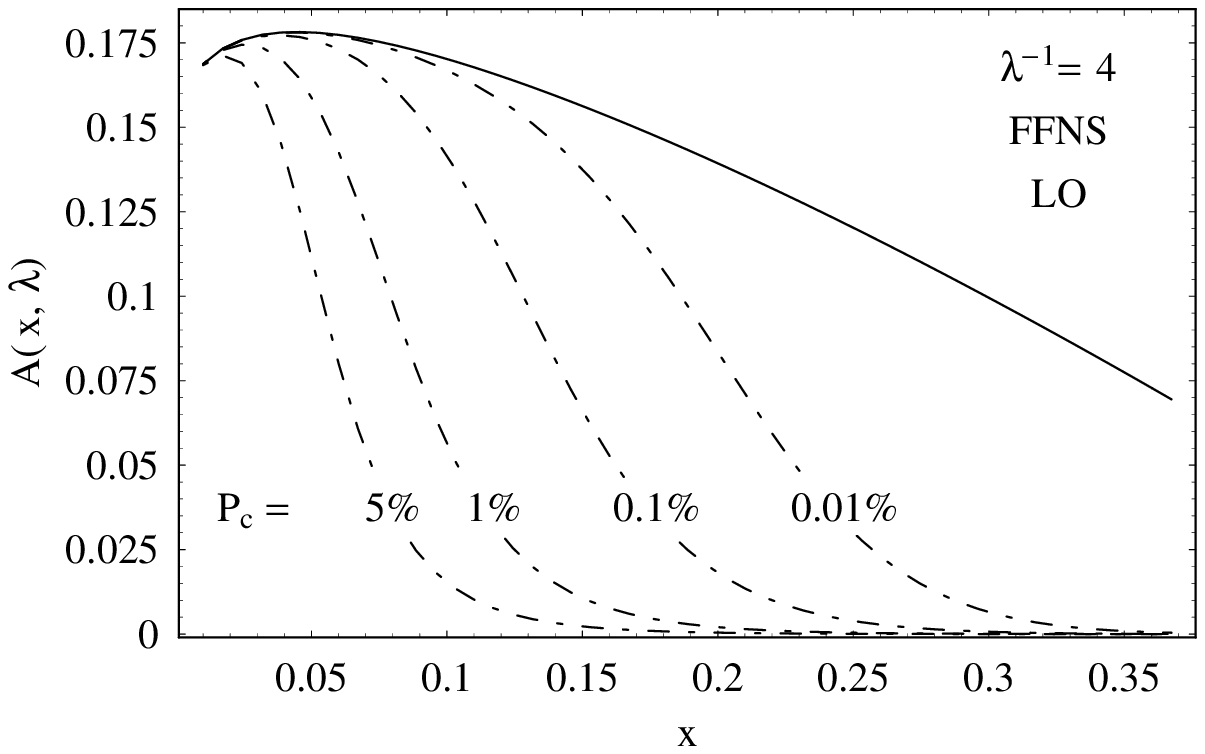,width=245pt}}\\
\mbox{\epsfig{file=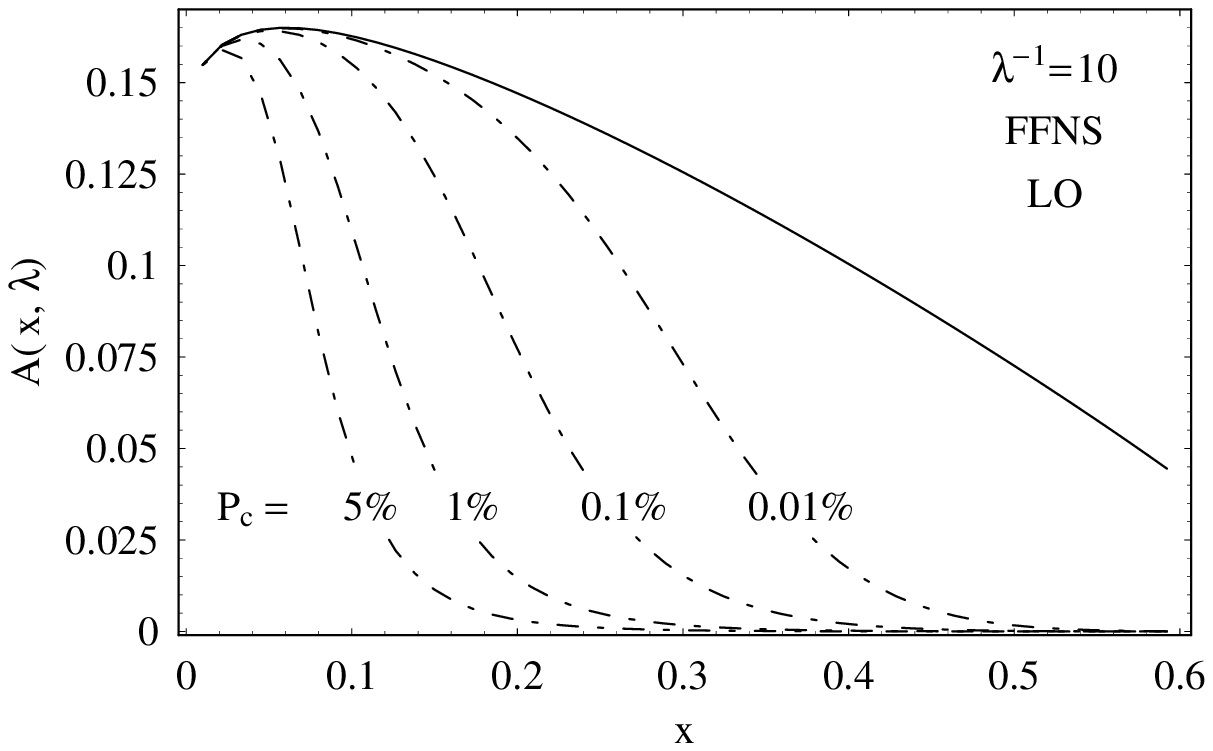,width=243pt}}
& \mbox{\epsfig{file=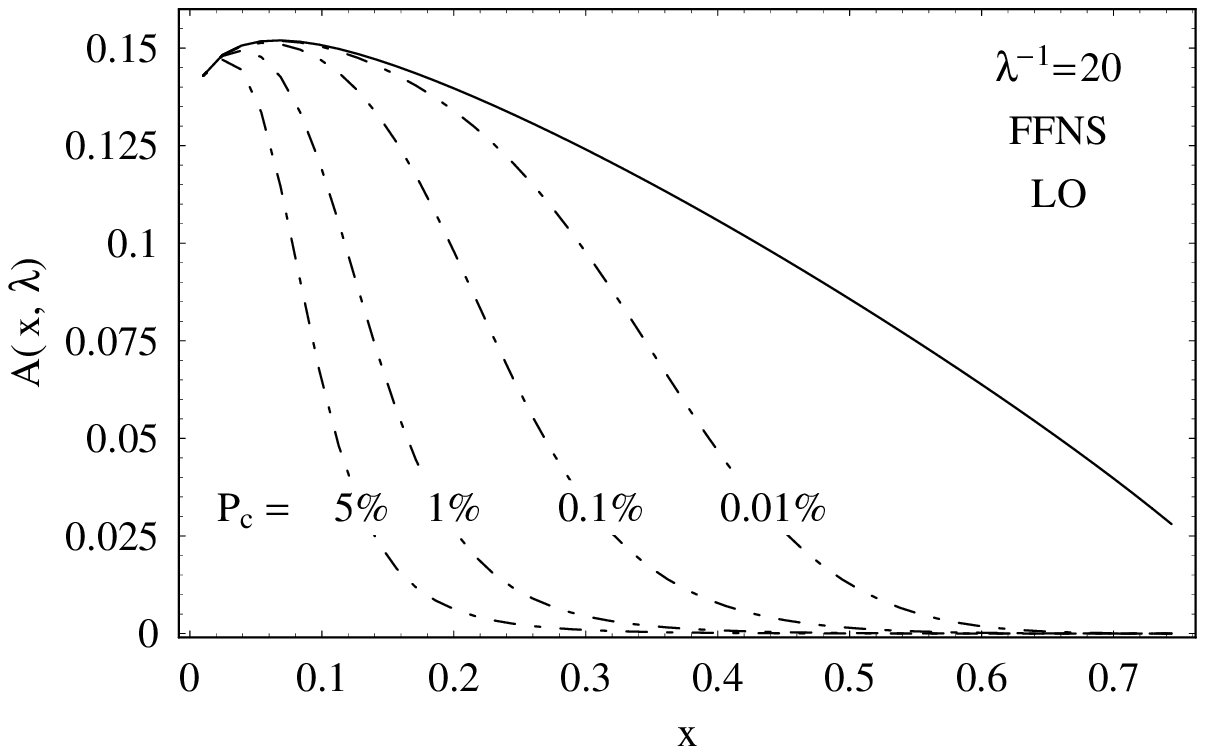,width=245pt}}\\
\mbox{\epsfig{file=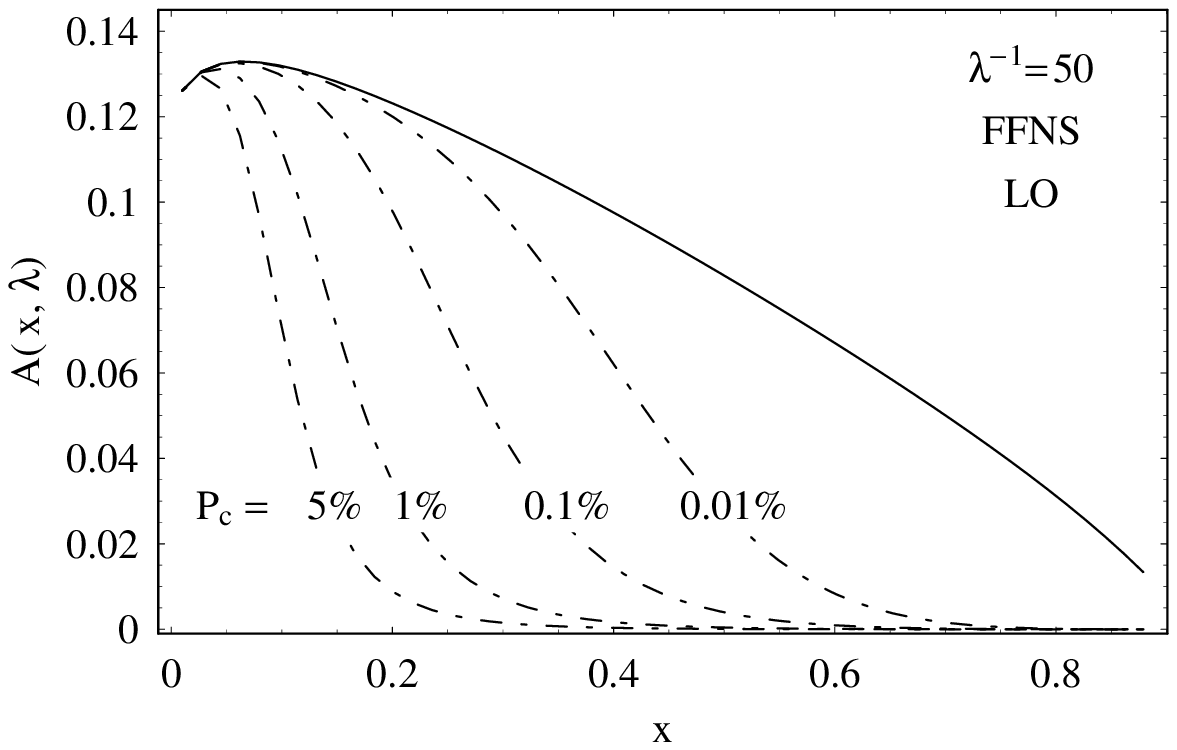,width=250pt}}
& \mbox{\epsfig{file=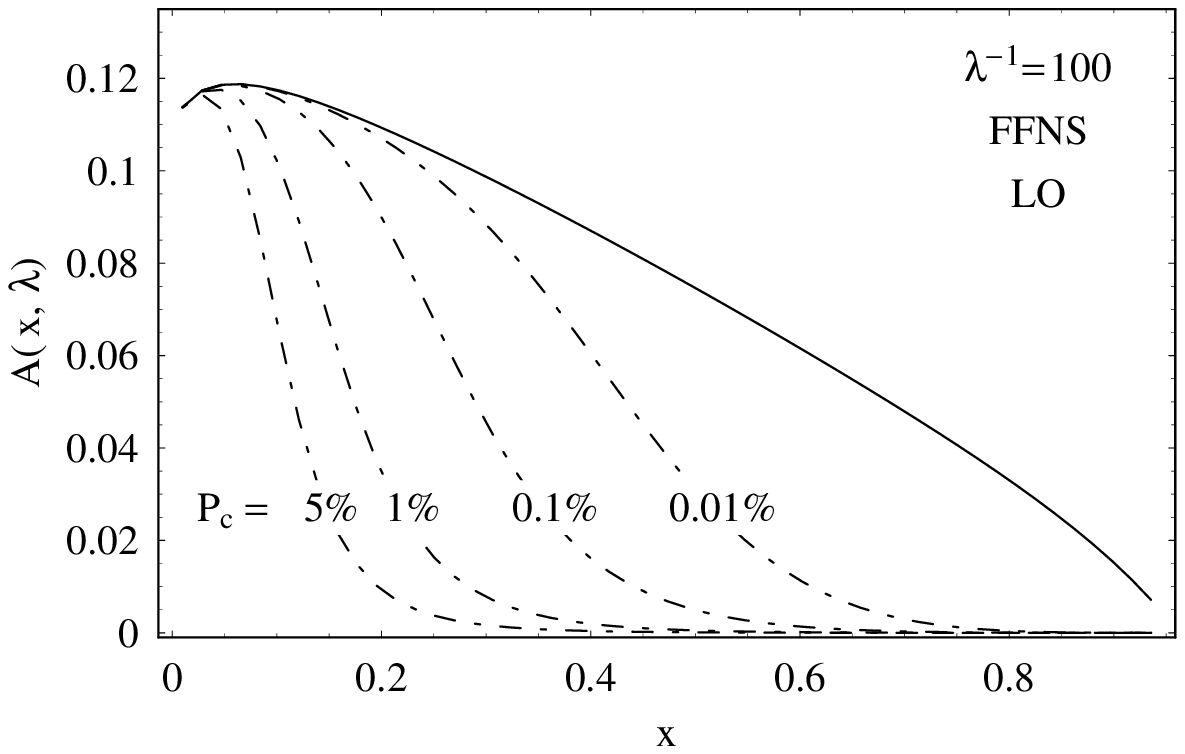,width=247pt}}\\
\end{tabular}
\caption{\label{Fg.6a}\small The LO predictions for $A(x,\lambda)$ in the FFNS at several values of
$\lambda$ and $P_{c}=\int^{1}_{0}c(z)\text{d}z$. Dash-dotted curves describe the
$\text{GF}^{\text{(LO)}}$+$\text{QS}^{\text{(LO)}}$ contributions with $P_{c}=5\%,~1\%,~0.1\%$ and
$0.01\%$. Solid lines correspond to the case when $P_{c}=0$.}
\end{center}
\end{figure}

In Fig.~\ref{Fg.6a}, the dependence of the asymmetry $A(x,\lambda)$ on the nonperturbative
intrinsic charm content of the proton is presented. We plot the LO predictions for $A(x,\lambda)$
as a function of $x$ for several values of the variable $\lambda$ and quantity
$P_{c}=\int^{1}_{0}c(z)\text{d}z$ describing a probability for IC in the nucleon. Dash-dotted
curves describe the $\text{GF}^{\text{(LO)}}$+$\text{QS}^{\text{(LO)}}$ contributions with
$P_{c}=5\%,~1\%,~0.1\%$ and $0.01\%$. Solid lines correspond to the case when $P_{c}=0$. Comparing
with each other Figs.~\ref{Fg.6} and \ref{Fg.6a}, one can see that even a $0.1\%$ contribution of
the nonperturbative IC to the proton wave function could be extracted from the $\cos 2\varphi $
asymmetry at large enough Bjorken $x$.

\subsection{\label{hb} Variable Flavor Number Scheme and Perturbative Intrinsic Charm}

One can see from Eqs.~(\ref{39}) that the GF cross section $\hat{\sigma}_{2,g}^{(0)}(z,\lambda)$
contains potentially large logarithm, $\ln (Q^{2}/m^{2})$. The same situation takes also place for
the QS cross section $\hat{\sigma}_{2,Q}^{(1)}(z,\lambda)$ given by Eq.~(\ref{25}). At high
energies, $Q^{2}\rightarrow \infty$, the terms of the form $\alpha_{s}\ln (Q^{2}/m^{2})$ dominate
the production cross sections. To improve the convergence of the perturbative series at high
energies, the so-called variable flavor number schemes (VFNS) have been proposed. Originally, this
approach was formulated by Aivazis, Collins, Olness and Tung (ACOT) \cite{AOT,ACOT}.

In the VFNS, the mass logarithms of the type $\alpha_s^n\ln^n (Q^{2}/m^{2})$ are resummed via the
renormalization group equations. In practice, the resummation procedure consists of two steps.
First, the mass logarithms have to be subtracted from the fixed order predictions for the partonic
cross sections in such a way that in the asymptotic limit $Q^{2}\rightarrow \infty$ the well known
massless $\overline{\text{MS}}$ coefficient functions are recovered. Instead, a charm parton
density in the hadron, $c(x,Q^{2})$, has to be introduced. This density obeys the usual massless
NLO DGLAP evolution equation with the boundary condition $c(x,Q^{2}=Q_{0}^2)=0$ where $Q_{0}^2\sim
m^{2}$. So, we may say that, within the VFNS, the charm density arises perturbatively from the
$g\rightarrow c\bar{c}$ evolution.

In the VFNS, the treatment of the charm depends on the values chosen for $Q^{2}$. At low
$Q^{2}<Q_{0}^2$, the production cross sections are described by the light parton contributions
($u,d,s$ and $g$). The charm production is dominated by the GF process and its higher order QCD
corrections. At high $Q^{2}\gg m^{2}$, the charm is treated in the same way as the other light
quarks and it is represented by a charm parton density in the hadron, which evolves in $Q^{2}$. In
the intermediate scale region, $Q^{2}\sim m^{2}$, one has to make a smooth connection between the
two different prescriptions.

Strictly speaking, the perturbative charm density is well defined at high $Q^2\gg m^2$ but does not
have a clean interpretation at low $Q^2$. Since the perturbative IC originates from resummation of
the mass logarithms of the type $\alpha_s^n\ln^n (Q^{2}/m^{2})$, it is usually assumed that the
corresponding PDF vanishes with these logarithms, i.e. for $Q^{2}<Q_{0}^2\approx m^{2}$. On the
other hand, the threshold constraint $W^2=(q+p)^2=Q^2(1/x-1)>4m^2$ implies that $Q_0$ is not a
constant but "live" function of $x$. To avoid this problem, several solutions have been proposed
(see e.g. Refs.~\cite{chi,SACOT}). In this paper, we use the so-called ACOT($\chi$) prescription
\cite{chi} which guarantees (at least at $Q^2>m^2$) the correct threshold behavior of the
heavy-quark-initiated contributions.

Within the VFNS, the $\varphi$-independent charm production cross sections have three pieces:
\begin{equation}\label{52}
\sigma_{2}(x,\lambda)=\sigma_{2,GF}(x,\lambda)-\sigma_{2,SUB}(x,\lambda)+\sigma_{2,QS}(x,\lambda),
\end{equation}
where the first and third terms on the right hand side describe the usual (unsubtracted) GF and QS
contributions while the second (subtraction) term renders the total result infra-red safe in the
limit $m^{2}\rightarrow 0$. The only constraint imposed on the subtraction term is to reproduce at
high energies the familiar $\overline{\text{MS}}$ partonic cross section:
\begin{equation}\label{53}
\lim_{\lambda\rightarrow 0}\left[\hat{\sigma}_{2,g}(z,\lambda)-
\hat{\sigma}_{2,SUB}(z,\lambda)\right]=\hat{\sigma}^{\overline{\text{MS}}}_{2,g}(z).
\end{equation}
Evidently, there is some freedom in the choice of finite mass terms of the form $\lambda^{n}$ (with
a positive $n$) in $\hat{\sigma}_{2,SUB}(z,\lambda)$. For this reason, several prescriptions have
been proposed to fix the subtraction term. As mentioned above, we use the so-called ACOT($\chi$)
scheme \cite{chi}.

According to the ACOT($\chi$) prescription, the lowest order $\varphi$-independent cross section is
\begin{equation}\label{54}
\sigma^{(LO)}_{2}(x,\lambda)=\int\limits_{\chi}^{1}\text{d}z\,g(z,\mu_{F})\left[\hat{\sigma}_{2,g}^{(0)}
\!\left(x/z,\lambda\right)-\frac{\alpha_{s}}{\pi}\ln\frac{\mu_{F}^{2}}{m^{2}}
\;\hat{\sigma}_{B}\left(x/z\right)P^{(0)}_{g\rightarrow
c}\left(\chi/z\right)\right]+\hat{\sigma}_{B}(x)c_{+}(\chi,\mu_{F}),
\end{equation}
where $P^{(0)}_{g\rightarrow c}$ is the LO gluon-quark splitting function, $P^{(0)}_{g\rightarrow
c}(\zeta)=\left.\left[(1-\zeta)^{2}+\zeta^{2}\right]\right/2$, and the LO GF cross section
$\hat{\sigma}_{2,g}^{(0)}$ is given by Eqs.~(\ref{39}). Remember also that $\chi=x(1+4\lambda)$ and
$c_{+}(\zeta,\mu_{F})=c(\zeta,\mu_{F})+\bar{c}(\zeta,\mu_{F})$.

The asymptotic behavior of the subtraction terms is fixed by the parton level factorization
theorem. This theorem implies that the partonic cross sections d$\hat{\sigma}$ can be factorized
into process-dependent infra-red safe hard scattering cross sections d$\tilde{\sigma}$, which are
finite in the limit $m\rightarrow 0$, and universal (process-independent) partonic PDFs
$f_{a\rightarrow i}$ and fragmentation functions $d_{n\rightarrow Q}$:
\begin{equation}\label{add1}
\text{d}\hat{\sigma}(\gamma^{*}+a\rightarrow Q+X)=\sum_{i,n}f_{a\rightarrow
i}(\zeta)\otimes\text{d}\tilde{\sigma}(\gamma^{*}+i\rightarrow n+X)\otimes d_{n\rightarrow Q}(z).
\end{equation}
In Eq.~(\ref{add1}), the symbol $\otimes$ denotes the usual convolution integral, the indices
$a,i,n$ and $Q$ denote partons, $p_{i}=\zeta p_{a}$ and $p_{Q}=z p_{n}$. All the logarithms of the
heavy-quark mass (i.e., the singularities in the limit $m\rightarrow 0$) are contained in the PDFs
$f_{a\rightarrow i}$ and fragmentation functions $d_{n\rightarrow Q}$ while d$\tilde{\sigma}$ are
IR-safe (i.e., are free of the $\ln m^{2}$ terms). The expansion of Eq.~(\ref{add1}) can be used to
determine order by order the subtraction terms. In particular, for the LO GF contribution to the
charm leptoproduction one finds \cite{ACOT}
\begin{equation}\label{add2}
\hat{\sigma}^{(0)}_{k,SUB}\left(z,\ln\,(\mu_{F}^{2}/m^{2})\right)=f^{(1)}_{g\rightarrow
c}\left(\zeta,\ln\,(\mu_{F}^{2}/m^{2})\right)\otimes\hat{\sigma}^{(0)}_{k,QS}(z/\zeta), \qquad
\qquad  (k=2,L,A,I),
\end{equation}
where $f^{(1)}_{g\rightarrow c}\left(\zeta,\ln\,(\mu_{F}^{2}/m^{2})\right)=\left(
\alpha_{s}/2\pi\right)\ln\,(\mu_{F}^{2}/m^{2})\,P^{(0)}_{g\rightarrow c}\left(\zeta\right)$
describes the charm distribution in the gluon within the $\overline{\text{MS}}$ factorization
scheme.

One can see from Eq.~(\ref{add2}) that the azimuth-dependent GF cross sections
$\hat{\sigma}_{A,GF}$ and $\hat{\sigma}_{I,GF}$ don't have subtraction terms at LO because the
lowest order QS contribution is $\varphi$-independent. For this reason, the
$\cos2\varphi$-dependence within the VFNS has the same form as in the FFNS:
\begin{equation}\label{55}
\sigma^{(LO)}_{A}(x,\lambda)=\int\limits_{\chi}^{1}\text{d}z\,g(z,\mu_{F})\,\hat{\sigma}_{A,g}^{(0)}
\!\left(x/z,\lambda\right).
\end{equation}
\begin{figure}
\begin{center}
\begin{tabular}{cc}
\mbox{\epsfig{file=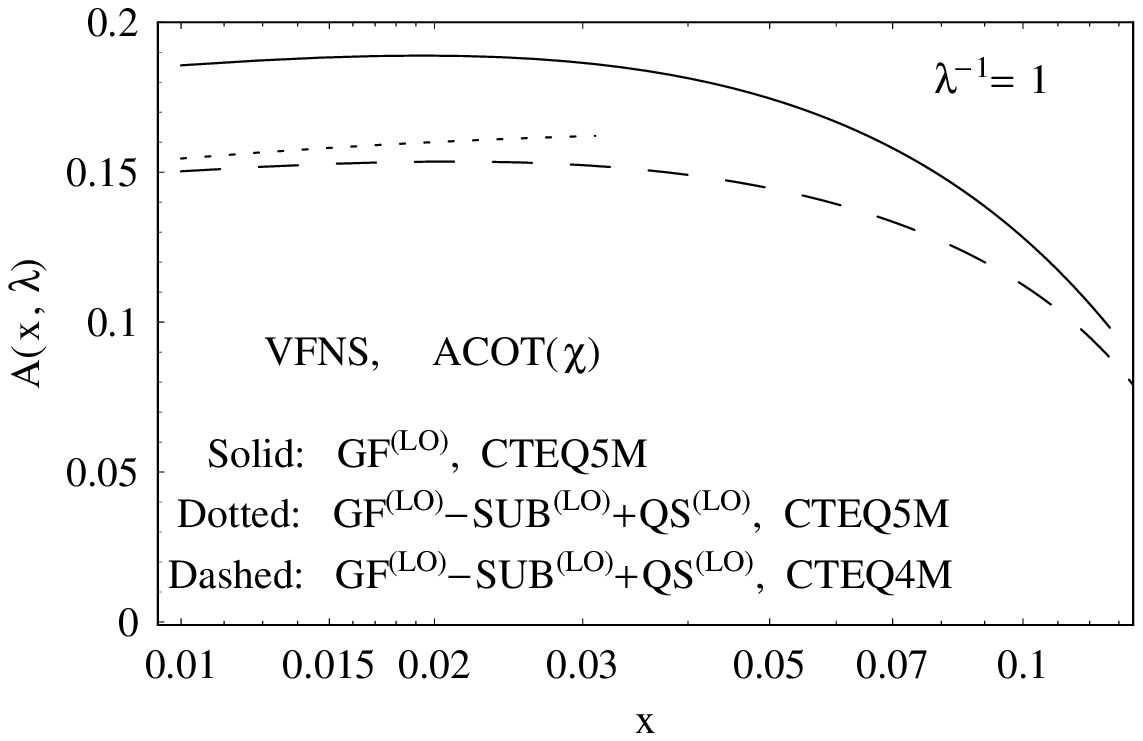,width=250pt}}
& \mbox{\epsfig{file=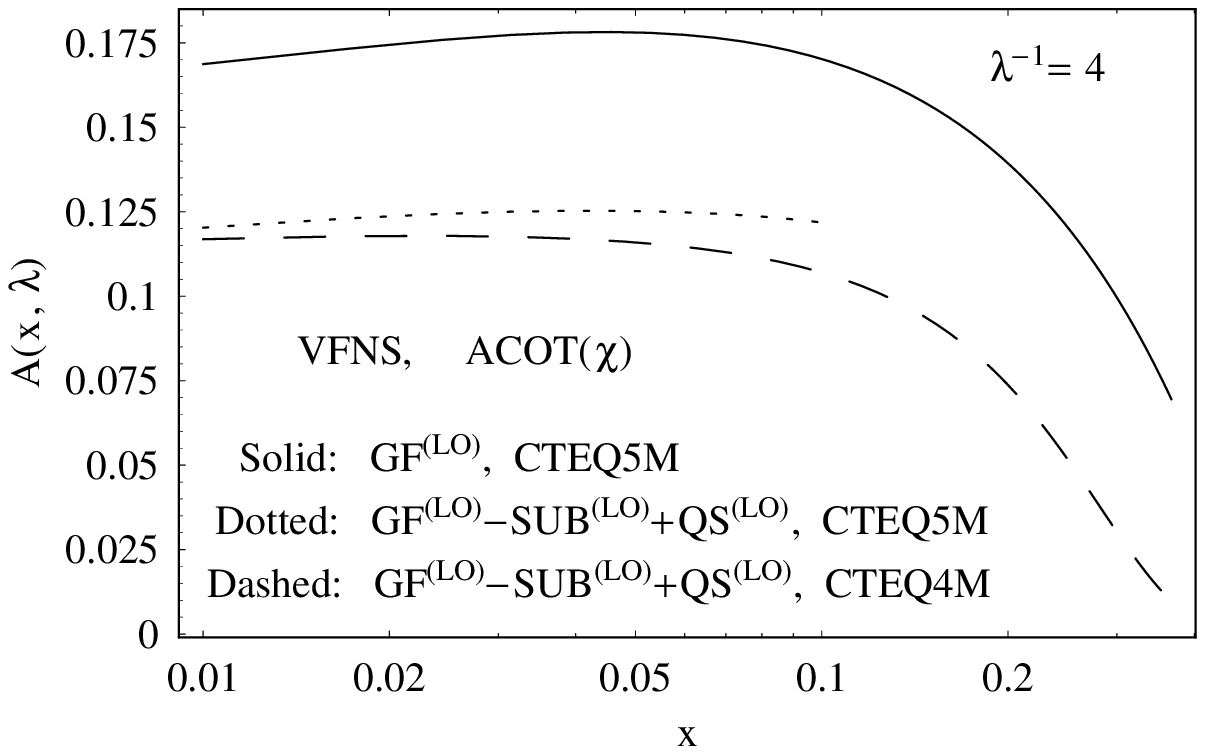,width=244pt}}\\
\mbox{\epsfig{file=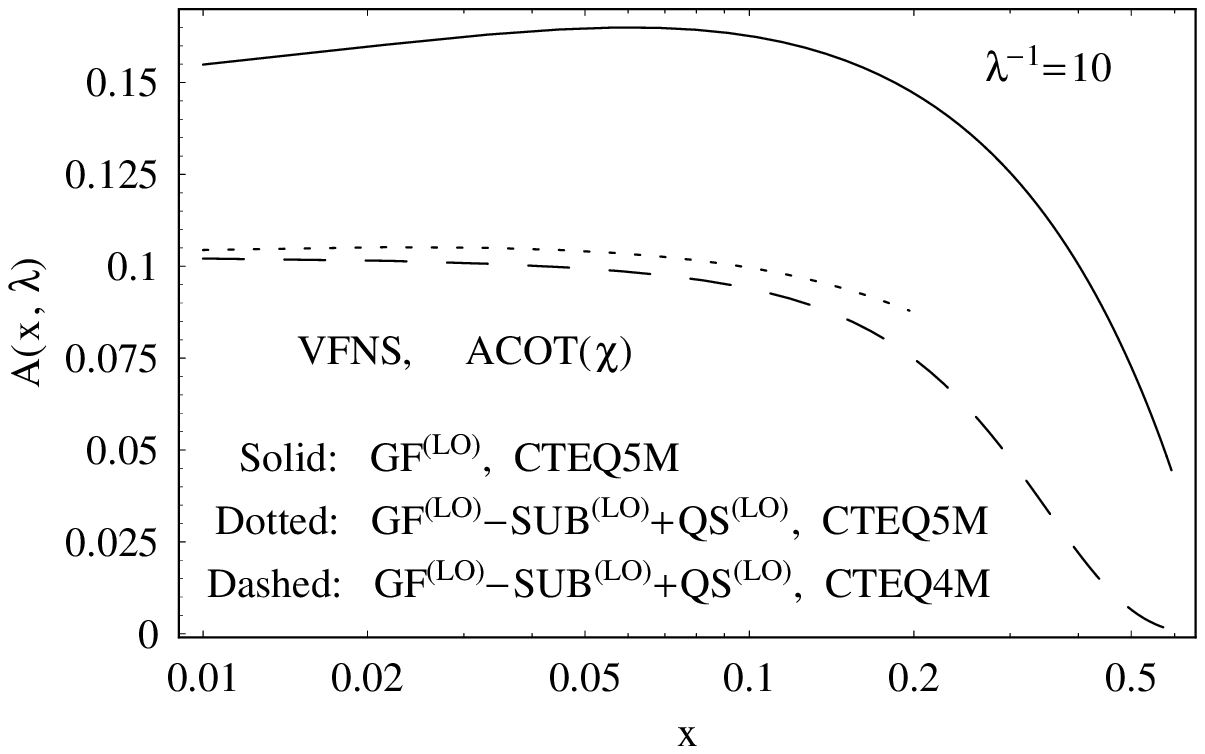,width=242pt}}
& \mbox{\epsfig{file=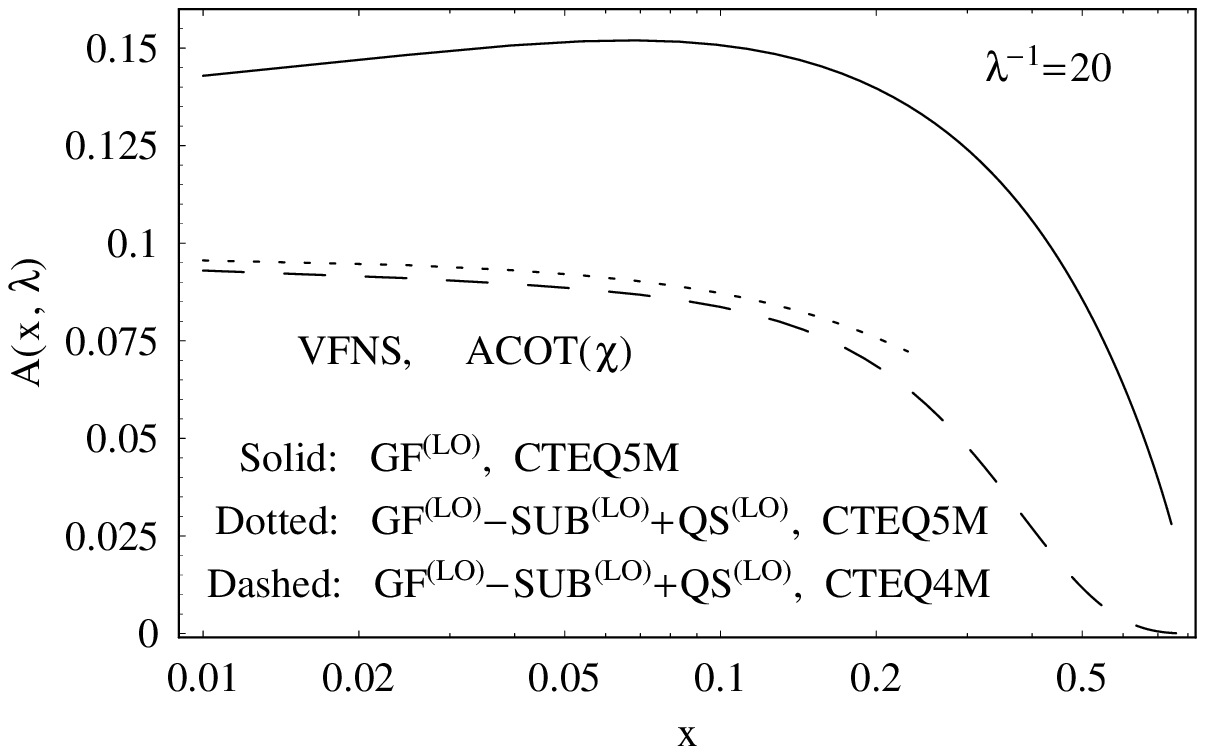,width=244pt}}\\
\mbox{\epsfig{file=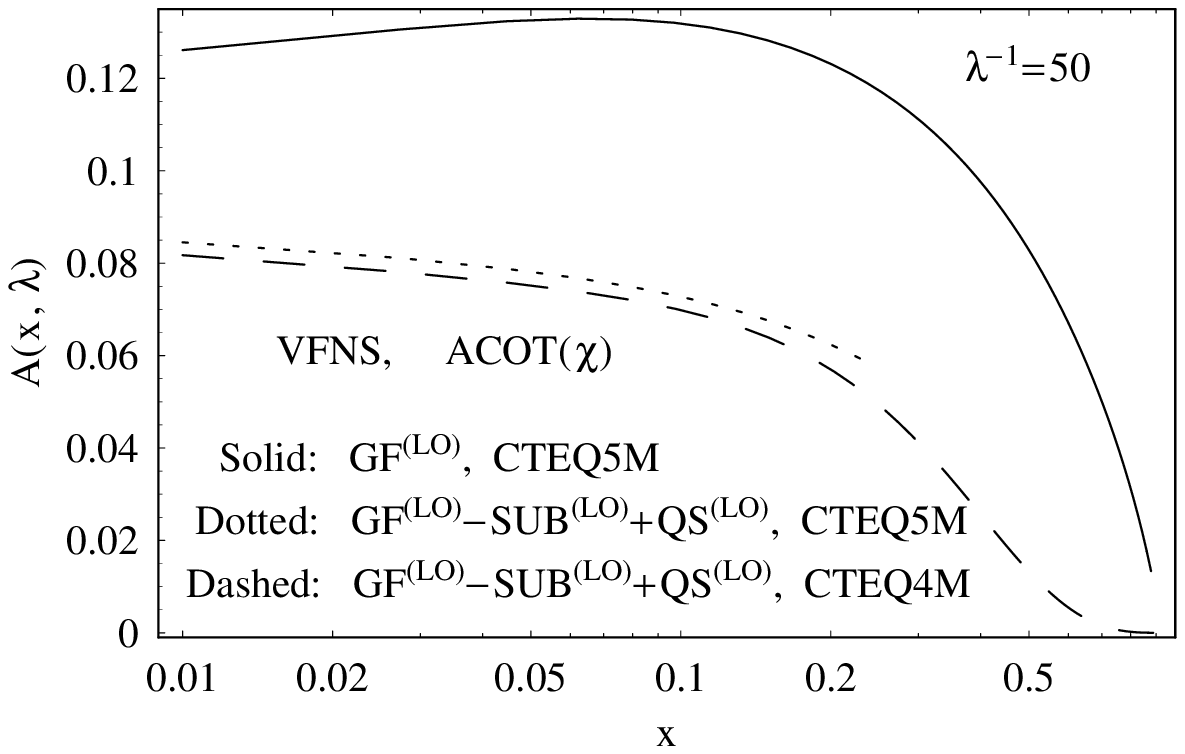,width=244pt}}
& \mbox{\epsfig{file=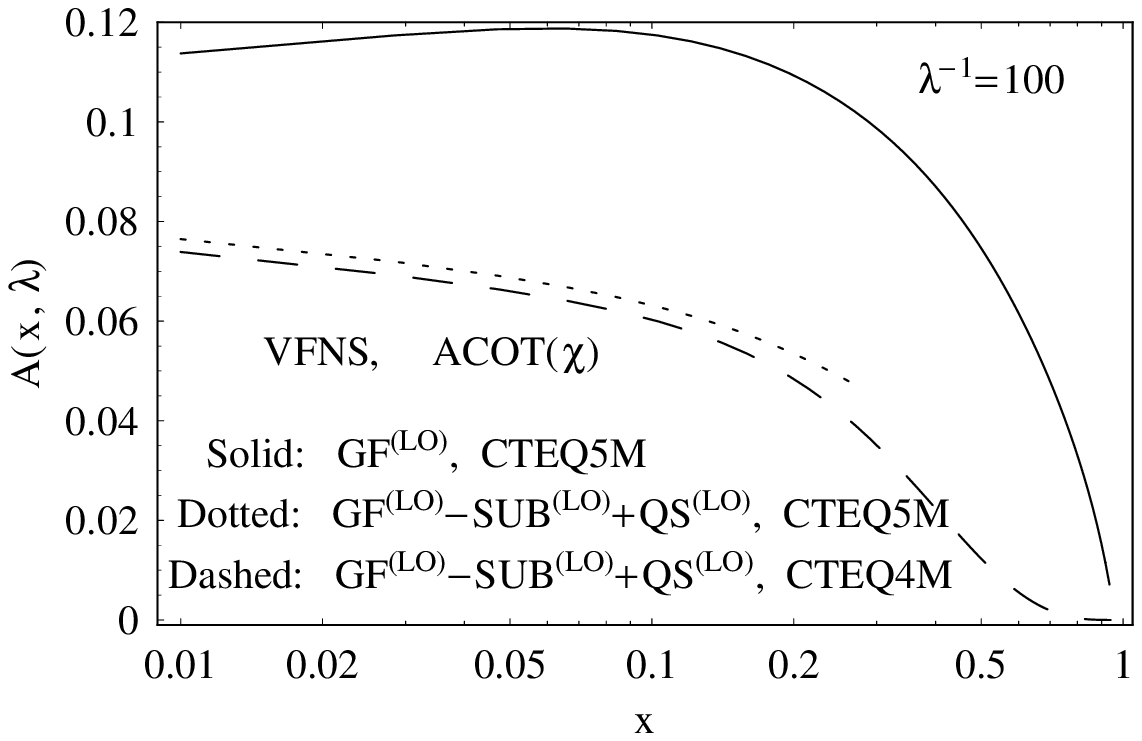,width=253pt}}\\
\end{tabular}
\caption{\label{Fg.7}\small Azimuthal asymmetry parameter $A(x,\lambda)$ in the VFNS at several
values of $\lambda$. The following contributions are plotted: $\text{GF}^{\text{(LO)}}$ (solid
curves), $\text{GF}^{\text{(LO)}}$$-\text{SUB}^{\text{(LO)}}$+$\text{QS}^{\text{(LO)}}$ with the
CTEQ5M set of PDFs (dotted curves) and $\text{GF}^{\text{(LO)}}$$-\text{SUB}^{\text{(LO)}}$+
$\text{QS}^{\text{(LO)}}$ with the CTEQ4M set of PDFs (dashed curves).}
\end{center}
\end{figure}

Fig.~\ref{Fg.7} shows the ACOT($\chi$) predictions for the asymmetry parameter $A(x,\lambda)$ at
several values of variable $\lambda$: $\lambda^{-1}=1,4,10,20,50$ and 100. For comparison, we plot
also the LO GF predictions (solid curves). In the ACOT($\chi$) case, we consider the CTEQ5M (dotted
lines) and CTEQ4M (dashed curves) parametrizations of the gluon and charm densities in the proton.
Corresponding values of the charm quark mass are $m_{c}=1.3 $ GeV \cite{CTEQ5} (for the CTEQ5M
PDFs) and $m_{c}=1.6 $ GeV \cite{CTEQ4} (for the CTEQ4M PDFs). The default value of the
factorization scale is $\mu_{F}=\sqrt{m^{2}+Q^{2}}$.

One can see from Fig.~\ref{Fg.7} the following properties of the azimuthal asymmetry,
$A(x,\lambda)$, within the VFNS. Contrary to the nonperturbative IC component, the perturbative one
is significant practically at all values of Bjorken $x$ and $Q^{2}>m^{2}$. The perturbative charm
contribution leads to a sizeable decreasing of the GF predictions for the $\cos2\varphi$-asymmetry.
In the ACOT($\chi$) scheme, the IC contribution reduces the GF results for $A(x,\lambda)$ by about
$30\%$. The origin of this reduction is straightforward: the QS component is practically
$\cos2\varphi$-independent.

The ACOT($\chi$) predictions for the asymmetry depend weakly on the parton distribution functions
we use. It is seen from Fig.~\ref{Fg.7} that the CTEQ5M and CTEQ4M sets of PDFs lead to very
similar results  for $A(x,\lambda)$. Note that we give the CTEQ5M predictions at low $x$ only
because of irregularities in the CTEQ5M charm density at large $x$.

We have also analyzed how the VFNS predictions depend on the choice of subtraction prescription. In
particular, the schemes proposed in Refs.~\cite{KS,SACOT} have been considered. We find that,
sufficiently above the production threshold, these subtraction prescriptions reduce the GF results
for the asymmetry by approximately $30\div 50 \%$.

One can conclude that impact of the perturbative IC on the $\cos2\varphi$ asymmetry is essential in
the whole region of Bjorken $x$ and therefore can be tested experimentally.

\section{Conclusion}
In the present paper, we consider the azimuthal dependence in charm leptoproduction as a probe of
the IC content of the proton. Our analysis is based on the fact that the GF and QS components have
strongly different $\cos2\varphi$-distributions. This fact follows from the NLO calculations of
both parton level contributions. In the framework of the FFNS, we justify the most remarkable
property of the hadron level azimuthal $\cos2\varphi$ asymmetry: the combined GF+QS predictions for
$A(x,Q^{2})$ are perturbatively and parametrically stable. The nonperturbative IC contribution
(resulting from the five-quark $\left\vert uudc\bar{c}\right\rangle$ component  of the proton wave
function) is practically invisible at low $x$, but affects essentially the GF predictions for the
asymmetry at large Bjorken $x$. We conclude that measurements of the $\cos2\varphi$ asymmetry at
large $x$ could directly probe the nonperturbative intrinsic charm.

Within the VFNS, charm density originates perturbatively from the $g\rightarrow c\bar{c}$ process
and obeys the DGLAP evolution equation. Presently, charm densities are included practically in all
the global sets of PDFs like CTEQ and MRST. Our analysis shows that these charm distribution
functions reduce dramatically (by about 1/3) the GF predictions for $A(x,Q^{2})$ practically at all
values of $x$. For this reason, the perturbative IC contribution can easily be measured using the
azimuthal $\cos2\varphi$-distributions in charm leptoproduction.

The VFN schemes have been proposed to resum the mass logarithms of the form $\alpha_{s}^{n}\ln^{n}
(Q^{2}/m^{2})$ which dominate the production cross sections at high energies. Evidently, were the
calculation done to all orders of $\alpha_{s}$, the VFNS and FFNS (without nonperturbative IC)
would be exactly equivalent. There is a point of view advocated in Refs.~\cite{ACOT,collins} that,
at high energies, the perturbative series converge better within the VFNS than in the FFNS. There
is also another opinion \cite{BMSN,neerven} that the above logarithms do not vitiate the
convergence of the perturbation expansion so that a resummation is, in principle, not necessary.
Our analysis of the azimuthal dependence in leptoproduction indicates an experimental way to
resolve this problem. First, contrary to the production cross sections, the azimuthal
$\cos2\varphi$-asymmetry is well defined numerically in pQCD. Second, sufficiently above the
production threshold (i.e., at small enough Bjorken $x$), the LO VFNS predictions for the
$\cos2\varphi$-asymmetry differ by more than $30\%$ from the corresponding FFNS ones. Third,
nonperturbative contributions (like the intrinsic gluon motion in the target) can't compensate for
this difference at non-small $Q^{2}$ where the VFNS is expected to be adequate. Therefore
measurements of the azimuthal distributions in charm leptoproduction would make it possible to
clarify the question whether the VFNS perturbative series for $A(x,Q^{2})$ converges better than
the FFNS one.
\begin{acknowledgments}
We thank S.J. Brodsky for stimulating discussions and useful suggestions. We also would like to
acknowledge interesting correspondence with I. Schienbein. This work was supported in part by
the ANSEF grant 04-PS-hepth-813-98 and NFSAT grant GRSP-16/06.
\end{acknowledgments}

\appendix
\section{\label{virt}Virtual and Soft Contributions to the Quark Scattering}
In this Appendix we reproduce some results of Hoffmann and Moore for the $\varphi$-independent QS
cross sections, and correct two misprints uncovered in Ref.~\cite{HM}. We work in four dimensions,
in the Feynman gauge and use the on-mass-shell renormalization scheme. We compute the absorptive
part of the Feynman diagram (which is free of the UV divergences) and then restore the real part
using the appropriate dispersion relations.

In the on-mass-shell scheme, the renormalized fermion self-energy vanishes like
$(\hat{p}_{Q}-m)^{2}$ which means that the second and third diagrams in Fig.~\ref{Fg.2}c do not
contribute to the cross section when the external quark legs are on-shell, $\hat{p}_{Q}\rightarrow
m$. The first graph in Fig.~\ref{Fg.2}c describes the NLO corrections to the quark-photon vertex
function:
\begin{equation}  \label{71}
\Lambda _{\mu }(q) =f\left( Q^{2}\right) \gamma _{\mu }- \frac{g\left( Q^{2}\right) }{2m}\sigma
_{\mu \nu }q^{\nu },
\end{equation}
where $\sigma _{\mu \nu }=\frac{1}{2}\left( \gamma _{\mu }\gamma _{\nu }-\gamma _{\nu }\gamma _{\mu
}\right) $ while $f\left( Q^{2}\right) $ and $g\left( Q^{2}\right)$ are the quark electromagnetic
formfactors. At the lowest order $\Lambda _{\mu }^{(0)}=\gamma _{\mu }$.

The virtual lepton-quark cross section, $\hat{\sigma}_{lQ}^{V}$, is obtained from the interference
term between the virtual and the Born amplitude. The result can be written in terms of the
electromagnetic formfactors as:
\begin{equation} \label{72}
\frac{\text{d}^{\text{2}}\hat{\sigma}_{lQ}^{V}}{\text{d}z\text{d}Q^{2}}= \frac{\alpha_{em}}{\pi
}\frac{\hat{\sigma}_{B}(z)}{zQ^{2}}\delta (1-z)\left\{ \left[ 1+(1-y)^{2}-2\lambda
z^{2}y^{2}\right] f\left( Q^{2}\right) +y^{2}g\left( Q^{2}\right) \right\}.
\end{equation}
Taking into account the definition of the HM cross sections, $\sigma^{(2)}$ and $\sigma^{(L)}$,
given by Eqs.~(\ref{35}), (\ref{36}) and (\ref{18}), we find that corresponding virtual parts are:
\begin{equation}\label{73}
\sigma_{1V}^{(2)}(z,Q^{2})=2\delta (1-z)f^{(1)}\left( Q^{2}\right), \qquad \qquad \qquad \sigma
_{1V}^{(L)}(z,Q^{2})=-\delta (1-z)g^{(1)}\left( Q^{2}\right),
\end{equation}
where $f^{(1)}\left( Q^{2}\right)$ and $g^{(1)}\left( Q^{2}\right)$ are the NLO corrections to the
electromagnetic formfactors. For the NLO HM  cross sections, $\sigma_{1V}^{(2)}$ and
$\sigma_{1V}^{(L)}$, we use exactly the same notations as in Ref.~\cite{HM}.

In the on-mass-shell renormalization scheme, the renormalized vertex correction vanishes as the
photon virtuality goes to zero, $f^{(1)}\left(0\right)=0$. This is a consequence of the Ward
identity and the fact that the real photon field (like the massive fermion one) is unrenormalized
in first order QCD. To satisfy the condition $f^{(1)}\left(0\right)=0$ automatically, we should use
for $f^{(1)}\left(q^{2}\right)$ the dispersion relation with one subtraction. The second
formfactor, $g^{(1)}\left(q^{2}\right)$, has no singularities. For this reason, we use for
$g^{(1)}\left(q^{2}\right)$ the dispersion relation without subtractions:
\begin{equation}\label{74}
f^{(1)}\left( q^{2}\right) =\frac{q^{2}}{\pi }\int\limits_{4m^{2}}^{\infty }
\frac{\text{d}t\;\,\text{Im }f^{(1)}(t)}{t(t-q^{2}-i0)},\qquad \qquad \qquad g^{(1)}\left(
q^{2}\right) =\frac{1}{\pi }\int\limits_{4m^{2}}^{\infty } \frac{\text{d}t\;\,\text{Im
}g^{(1)}(t)}{t-q^{2}-i0}.
\end{equation}
Calculating the imaginary parts of the formafactors and restoring their real parts with the help of
Eqs.~(\ref{74}) yields
\begin{eqnarray}
f^{(1)}\left( Q^{2}\right)&=&\frac{\alpha _{s}}{\pi }C_{F}\biggl\{ \left[ 1+ \frac{1+2\lambda
}{\sqrt{1+4\lambda }}\ln r\right]\left(\ln \frac{m}{m_{g}}-1\right) \nonumber\\%
&&+\frac{1+2\lambda}{\sqrt{1+4\lambda }}\left[ \text{Li}_{2}(-r)+ \frac{\pi ^{2}}{12}+
\frac{1}{4}\ln^{2}r+\frac{1}{2}\ln r\ln\frac{1+4\lambda }{\lambda }\right]+\frac{1}{4}
\frac{\ln r}{\sqrt{1+4\lambda }}\biggr\},\label{75}\\
g^{(1)}\left( Q^{2}\right)&=&-\frac{\alpha_{s}}{\pi}C_{F}\frac{\lambda\ln
r}{\sqrt{1+4\lambda}}.\label{76}
\end{eqnarray}
In Eqs.~(\ref{75}) and (\ref{76}), $C_{F}=(N_{c}^{2}-1)/(2N_{c})$, where $N_{c}$ is number of
colors, and $r$ is defined by Eq.~(\ref{29}). Taking into account that $\Lambda
^{(1)\mu}=\left(\alpha_{s}\left/\alpha_{em}\right.\right)C_{F}\,\Lambda_{QED}^{(1)\mu}$, we see
that Eqs.~(\ref{75}, \ref{76}) reproduce the textbook QED results.

It is now straightforward to obtain the virtual contribution to the longitudinal cross section.
Combining Eqs.~(\ref{73}) and (\ref{76}) yields:
\begin{equation}\label{77}
\sigma _{1V}^{(L)}(z,Q^{2})=\frac{\alpha _{s}}{\pi }C_{F}\delta (1-z)\frac{\lambda \ln
r}{\sqrt{1+4\lambda }}.
\end{equation}
Comparing the above result with the corresponding one given by Eq.~(39) in Ref.~\cite{HM}, we see
that the HM expression for $\sigma _{1V}^{(L)}$ has opposite sign. Note also that this typo
propagates into the final result for $\sigma _{1}^{(L)}$ given by Eq.~(52) \cite{HM}.

Calculation of the bremsstrahlung contribution to the longitudinal cross section, $\sigma
_{1B}^{(L)}(z,Q^{2})$, is also straightforward. We coincide with the HM result for $\sigma
_{1B}^{(L)}(z,Q^{2})$ given by Eq.~(49) in Ref.~\cite{HM}. However there is one more misprint in
the HM expression for $\sigma _{1}^{(L)}$: the r.h.s of  Eq.~(52) \cite{HM} should be multiplied by
$z$.

In the case of $\sigma _{1}^{(2)}(z,Q^{2})$, the situation is slightly more complicated due to the
need to take into account the IR singularities. One can see from Eq.~(\ref{75}) that $f^{(1)}\left(
Q^{2}\right)$ has an IR divergence which is regularized with the help of an infinitesimal gluon
mass $m_{g}$. This singularity is cancelled when one adds the so-called soft contribution
originating from the real gluon emission. For this purpose we introduce another infinitesimal
parameter $\delta z$, $\left(m_{g}\!\left/m\right.\right)\ll\delta z\ll 1$. The full bremsstrahlung
contribution, $\sigma _{1B}^{(2)}$, can then be splitted into the soft and hard pieces as follows:
\begin{equation}\label{78}
\sigma _{1soft}^{(2)}(z,Q^{2})=\theta (z+\delta z-1)\sigma _{1B}^{(2)}(z,Q^{2}),\qquad \qquad
\qquad \qquad \sigma _{1hard}^{(2)}(z,Q^{2})=\theta (1-z-\delta z)\sigma _{1B}^{(2)}(z,Q^{2}),
\end{equation}
where $\theta (1-z-\delta z)$ is the Heaviside step function. The soft cross section should be
calculated in the eikonal approximation, $\vec{p_{g}}\rightarrow 0$, taking into account the
infinitesimal gluon mass $m_{g}$. As a result, the sum of the virtual and soft contributions is IR
finite:
\begin{eqnarray}
\sigma _{1V}^{(2)}+\sigma _{1soft}^{(2)} &=&\frac{\alpha _{s}}{\pi } C_{F}\delta (1-z)\biggl\{-2\ln
(\delta z)\left[ 1+\frac{1+2\lambda }{\sqrt{ 1+4\lambda }}\ln r\right] +2\ln \lambda
-1-\frac{\sqrt{1+4\lambda }}{2}\ln r \nonumber\\
&&+\frac{1+2\lambda}{\sqrt{1+4\lambda}}\left[\text{Li}_{2}(r^{2})+2\text{Li}_{2}(-r)+\frac{3}{2}
\ln^{2}r-2\ln r-\ln r\ln\lambda+2\ln r\ln (1+4\lambda)\right] \biggr\}. \label{79}
\end{eqnarray}
Adding to the above expression the hard cross section $\sigma _{1hard}^{(2)}$ defined by
Eq.~(\ref{78}), we reproduce  in the limit $\delta z\rightarrow 0$ the full result for $\sigma
_{1}^{(2)}$ given by Eq.~(51) in Ref.~\cite{HM}.

\section{\label{soft}NLO Soft-Gluon Corrections to the Photon-Gluon Fusion}
This Appendix provides an overview of the NLO soft-gluon approximation for the photon-gluon fusion
mechanism. We present the final results for the parton level cross sections to the next-to-leading
logarithmic (NLL) accuracy. More details can be found in Refs.~\cite{Laenen-Moch,we2,we4}.

To take into account the NLO contributions to the GF mechanism, one needs to calculate the virtual
${\cal O}(\alpha _{em}\alpha _{s}^{2})$ corrections to the Born process (\ref{38}) and the real
gluon emission:
\begin{equation}  \label{60}
\gamma ^{*}(q)+g(k_{g})\rightarrow Q(p_{Q})+\overline{Q}(p_{\stackrel{\_}{Q}})+g(p_{g}).
\end{equation}
The partonic invariants describing the single-particle inclusive (1PI) kinematics are
\begin{eqnarray}
s^{\prime }=2q\cdot k_{g}=s+Q^{2}=\zeta S^{\prime },\qquad \qquad &&t_{1}=\left(
k_{g}-p_{Q}\right) ^{2}-m^{2}=\zeta T_{1},  \nonumber \\
s_{4}=s^{\prime }+t_{1}+u_{1},\qquad \qquad &&u_{1}=\left( q-p_{Q}\right) ^{2}-m^{2}=U_{1},
\label{61}
\end{eqnarray}
where $\zeta$ is defined by $\vec{k}_{g}= \zeta\vec{p}\,$ and $s_{4}$ measures the inelasticity of
the reaction (\ref{60}). The corresponding 1PI hadron level variables describing the reaction
(\ref{1}) are
\begin{eqnarray}
S^{\prime }=2q\cdot p=S+Q^{2},\qquad \qquad &&T_{1}=\left( p-p_{Q}\right)
^{2}-m^{2},  \nonumber \\
S_{4}=S^{\prime }+T_{1}+U_{1},\qquad \qquad &&U_{1}=\left( q-p_{Q}\right) ^{2}-m^{2}.  \label{62}
\end{eqnarray}

The exact NLO calculations of the unpolarized heavy quark production in $\gamma g$
\cite{Ellis-Nason,Smith-Neerven}, $\gamma ^{*}g$ \cite{LRSN}, and $gg$
\cite{Nason-D-E-1,Nason-D-E-2,Nason-D-E-3,BKNS} collisions show that, near the partonic threshold,
a strong logarithmic enhancement of the cross sections takes place in the collinear, $\vec{p}_{g,T}
$ $\rightarrow 0$, and soft, $\vec{p}_{g}\rightarrow 0$, limits. This threshold (or soft-gluon)
enhancement has universal nature in the perturbation theory and originates from incomplete
cancellation of the soft and collinear singularities between the loop and the bremsstrahlung
contributions. Large leading and next-to-leading threshold logarithms can be resummed to all orders
of perturbative expansion using the appropriate evolution equations
\cite{Contopanagos-L-S,Laenen-O-S,Kidonakis-O-S}. The analytic results for the resummed cross
sections are ill-defined due to the Landau pole in the coupling strength $\alpha _{s}$. However, if
one considers the obtained expressions as generating functionals of the perturbative theory and
re-expands them at fixed order in $\alpha _{s}$, no divergences associated with the Landau pole are
encountered.

Soft-gluon resummation for the photon-gluon fusion has been performed in Ref.~\cite{Laenen-Moch}
and checked in Refs.~\cite{we2,we4}. To NLL accuracy, the perturbative expansion for the partonic
cross sections, d$^{2}\hat{\sigma}_{k,g}/$d$t_{1}$d$u_{1}$ ($k=T,L,A,I$), can be written in a
factorized form as
\begin{equation}  \label{63}
s^{\prime 2}\frac{\text{d}^{2}\hat{\sigma}_{k,g}}{\text{d}t_{1}\text{d}u_{1}}( s^{\prime
},t_{1},u_{1}) =B_{k,g}^{\text{{\rm Born}}}( s^{\prime },t_{1},u_{1})\left\{\delta (s^{\prime
}+t_{1}+u_{1}) +\sum_{n=1}^{\infty }\left( \frac{\alpha _{s}C_{A}}{\pi}\right)^{n}K^{(n)}(
s^{\prime },t_{1},u_{1})\right\} ,
\end{equation}
with the Born level distributions $B_{k,g}^{\text{{\rm Born}}}$ given by
\begin{eqnarray}
B_{T,g}^{\text{{\rm Born}}}( s^{\prime },t_{1},u_{1}) &=&\pi e_{Q}^{2}\alpha _{em}\alpha _{s}\left[
\frac{t_{1}}{u_{1}}+\frac{u_{1}}{t_{1} }+4\left( \frac{s}{s^{\prime }}-\frac{m^{2}s^{\prime
}}{t_{1}u_{1}}\right) \left( \frac{s^{\prime }(m^{2}-Q^{2}/2)}{t_{1}u_{1}}+\frac{Q^{2}}{s^{\prime}
}\right) \right] ,  \nonumber \\
B_{L,g}^{\text{{\rm Born}}}( s^{\prime },t_{1},u_{1}) &=&\pi e_{Q}^{2}\alpha _{em}\alpha _{s}\left[
\frac{8Q^{2}}{s^{\prime }}\left( \frac{s}{s^{\prime }}-\frac{m^{2}s^{\prime }}{t_{1}u_{1}}\right)
\right] ,
\nonumber \\
B_{A,g}^{\text{{\rm Born}}}( s^{\prime },t_{1},u_{1}) &=&\pi e_{Q}^{2}\alpha _{em}\alpha _{s}\left[
4\left( \frac{s}{s^{\prime }}-\frac{ m^{2}s^{\prime }}{t_{1}u_{1}}\right) \left(
\frac{m^{2}s^{\prime }}{
t_{1}u_{1}}+\frac{Q^{2}}{s^{\prime }}\right) \right] ,  \label{64} \\
B_{I,g}^{\text{{\rm Born}}}( s^{\prime },t_{1},u_{1}) &=&\pi e_{Q}^{2}\alpha _{em}\alpha _{s}\left[
4\sqrt{Q^{2}}\left( \frac{t_{1}u_{1}s }{s^{\prime 2}}-m^{2}\right)
^{1/2}\frac{t_{1}-u_{1}}{t_{1}u_{1}}\left( 1-\frac{2Q^{2}}{s^{\prime }}-\frac{2m^{2}s^{\prime
}}{t_{1}u_{1}}\right) \right] .  \nonumber
\end{eqnarray}

Note that the functions $K^{(n)}( s^{\prime },t_{1},u_{1}) $ in Eq.~(\ref{63}) originate from the
collinear and soft limits. Radiation of soft and collinear gluons does not affect the transverse
momentum of detected particles and therefore the azimuthal angle $\varphi$. For this reason, the
functions $ K^{(n)}(s^{\prime },t_{1},u_{1}) $ are the same for all helicity cross sections
$\hat{\sigma}_{k,g}$ ($k=T,L,A,I$). At NLO, the soft-gluon corrections to NLL accuracy in the
$\overline{\text{MS}}$ scheme are
\begin{eqnarray}
K^{(1)}( s^{\prime },t_{1},u_{1}) &=&2\left[ \frac{\ln \left( s_{4}/m^{2}\right) }{s_{4}}\right]
_{+}-\left[ \frac{1}{s_{4}}\right] _{+}\left\{ 1+\ln \left( \frac{u_{1}}{t_{1}}\right) -\left(
1-\frac{2C_{F}}{ C_{A}}\right) \left( 1+\text{Re}L_{\beta }\right) +\ln \left( \frac{\mu ^{2}
}{m^{2}}\right) \right\}   \nonumber \\
&&+\delta ( s_{4}) \ln \left( \frac{-u_{1}}{m^{2}}\right) \ln \left( \frac{\mu ^{2}}{m^{2}}\right),
\label{65}
\end{eqnarray}
where we use $\mu =\mu _{F}=\mu_{R}$. In Eq.~(\ref{65}), $C_{A}=N_{c}$ and $
C_{F}=(N_{c}^{2}-1)/(2N_{c})$, where $N_{c}$ is number of colors, while $ L_{\beta
}=(1-2m^{2}/s)\{\ln [(1-\beta )/(1+\beta )]+$i$\pi\}$ with $\beta=\sqrt{1-4m^{2}/s}$. The
single-particle inclusive ''plus`` distributions are defined by
\begin{equation}  \label{66}
\left[\frac{\ln^{l}\left( s_{4}/m^{2}\right) }{s_{4}}\right]_{+}=\lim_{\epsilon \rightarrow
0}\left\{\frac{\ln^{l}\left(s_{4}/m^{2}\right) }{s_{4}}\theta ( s_{4}-\epsilon)+\frac{1}{l+1}\ln
^{l+1}\left(\frac{\epsilon }{m^{2}}\right) \delta ( s_{4})\right\}.
\end{equation}
For any sufficiently regular test function $h(s_{4})$, Eq.~(\ref{66}) gives
\begin{equation}\label{67}
\int\limits_{0}^{s_{4}^{\max }}\text{d}s_{4}\,h(s_{4})\left[ \frac{\ln ^{l}\left(
s_{4}/m^{2}\right) }{s_{4}}\right] _{+}=\int\limits_{0}^{s_{4}^{\max }}\text{d}s_{4}\left[
h(s_{4})-h(0)\right] \frac{\ln ^{l}\left( s_{4}/m^{2}\right) }{s_{4}}+\frac{1}{l+1}h(0)\ln
^{l+1}\left( s_{4}^{\max }/m^{2}\right) .
\end{equation}

In Eq.~(\ref{65}), we have preserved the NLL terms for the scale-dependent logarithms too. Note
also that the results (\ref{64}) and (\ref{65}) agree to NLL accuracy with the exact ${\cal
O}(\alpha_{em}\alpha _{s}^{2})$ calculations of the photon-gluon cross sections
$\hat{\sigma}_{T,g}$ and $\hat{\sigma}_{L,g}$ given in Ref.~\cite{LRSN}.

To investigate the scale dependence of the results (\ref{63}$-$\ref{65}), it is convenient to
introduce for the fully inclusive (integrated over $t_{1}$ and $u_{1}$) cross sections,
$\hat{\sigma}_{k,g}$ ($k=T,L,A,I$),{\large \ }the dimensionless coefficient functions
$c_{k,g}^{(n,l)}$ defined by Eq.~(\ref{32}). Concerning the NLO scale-independent coefficient
functions, only $ c_{T,g}^{(1,0)}$ and $c_{L,g}^{(1,0)}$ are known exactly
\cite{LRSN,Harris-Smith}. As to the $\mu$-dependent coefficients, they can by calculated explicitly
using the evolution equation:
\begin{equation}  \label{68}
\frac{\text{d}\hat{\sigma}_{k,g}(z ,Q^{2},\mu ^{2})}{\text{d}\ln \mu
^{2}}=-\int\limits_{\zeta_{\min }}^{1}\text{d}\zeta\,\hat{\sigma}_{k,g}(z/\zeta,Q^{2},\mu
^{2})P_{gg}(\zeta),
\end{equation}
where $z=Q^{2}/s^{\prime}$, $\zeta_{\min }=z(1+4\lambda)$, $\hat{\sigma}_{k,g}(z,Q^{2},\mu )$ are
the cross sections resummed to all orders in $\alpha _{s}$ and $P_{gg}(\zeta)$ is the corresponding
(resummed) Altarelli-Parisi gluon-gluon splitting function. Expanding Eq.~(\ref{68}) in $\alpha
_{s}$, one can find \cite{Laenen-Moch,we2}
\begin{equation}  \label{69}
c_{k,g}^{(1,1)}(z, \lambda)=\frac{1}{4\pi ^{2}}\int\limits_{\zeta_{\min}}^{1}\text{d}\zeta\left[
b_{2}\delta (1-\zeta)-\,P_{gg}^{(0)}(\zeta)\right] c_{k,g}^{(0,0)}(z/\zeta,\lambda),
\end{equation}
where $b_{2}=(11C_{A}-2n_{f})/12$ is the first coefficient of the $\beta(\alpha _{s})$-function
expansion and $n_{f}$ is the number of active quark flavors. The one-loop gluon splitting function
is:
\begin{equation}  \label{70}
P_{gg}^{(0)}(\zeta)=\lim_{\epsilon \rightarrow 0}\left\{ \left( \frac{\zeta}{1-\zeta}+
\frac{1-\zeta}{\zeta}+\zeta(1-\zeta)\right) \theta(1-\zeta-\epsilon )+\delta(1-\zeta)\ln \epsilon
\right\} C_{A}+b_{2}\delta(1-\zeta).
\end{equation}

With Eq.~(\ref{69}) in hand, it is possible to check the quality of the NLL approximation against
exact answers. As shown in Ref.~\cite{we4}, the soft-gluon corrections reproduce satisfactorily the
threshold behavior of the available exact results for $\lambda\sim1$. Since the gluon distribution
function supports just the threshold region, the soft-gluon contribution dominates the
photon-hadron cross sections $\sigma_{k,GF}$ ($k=T,L,A,I$) at energies not so far from the
production threshold and at relatively low virtuality $Q^{2}\lesssim m^{2}$.

\section{\label{exp}Nonperturbative IC and Relevant Experimental Facts}
The most clean probe of the charm quark distribution function (both perturbative and
nonperturbative) is the semi-inclusive deep inelastic lepton-proton scattering, $lp\rightarrow
l'cX$. To measure the nonperturbative IC contribution, one needs data on the charm production at
sufficiently large Bjorken $x$. The only experiment which has investigated the large $x$ domain is
the European Muon Collaboration (EMC) \cite{EMC} where the decay lepton spectra have been used to
detect the produced charmed particles. In Ref.~\cite{harris-emc}, a re-analysis of the EMC data on
$F_{2}^{c}(x,Q^{2})$ have been performed using the NLO results for both GF and QS components. The
analysis \cite{harris-emc} shows that a nonperturbative intrinsic charm contribution to the proton
wave function of the order of $1\%$ is needed to fit the EMC data in the large $x$ region. This
value of the nonperturbative IC is consistent with the estimates based on the operator product
expansion \cite{polyakov}. Note however that the EMC data are of limited statistics and, for this
reason, more accurate measurements of charm leptoproduction at large $x$ are necessary.

It is also possible to extract useful information on the IC from diffractive dissociation processes
such as $p\rightarrow p J/\psi$ on a nuclear target. Comprehensive measurements of the
$pA\rightarrow J/\psi X$ and $\pi A \rightarrow J/\psi X$ cross sections have been performed in the
fixed target experiments NA3 at CERN \cite{NA3} and E886 at FNAL \cite{E886}. According to the
arguments presented in Refs.~\cite{brod-psi-1,hoyer-psi-1,brod-psi-2}, the IC contribution is
predicted to be strongly shadowed in the above reactions that is in a complete agreement with the
observed nuclear dependence of the high Feynman $x_{F}$ component of the $J/\psi$ hadroproduction.

A non-vanishing five-quark Fock component $\left\vert uudc\bar{c}\right\rangle$ leads to the
production of open charm states such as $\Lambda_{c}(cud)$ and $D^{-}(\bar{c}d)$ with large Feynman
$x_{F}$. This may occur either through a coalescence of the valence and charm quarks which are
moving with the same rapidity or via hadronization of the produced $c$ and $\bar{c}$. As shown in
Refs.~\cite{barger-lead,brod-lead}, a model based on the nonperturbative intrinsic charm naturally
explains the leading particle effect in the $pp\rightarrow DX$ and $pp\rightarrow \Lambda_{c}X$
processes that has been observed at the ISR \cite{ISR-lead} and Fermilab
\cite{E791-lead-1,E791-lead-2}.

As to the high-$x_{F}$ hadroproduction of open bottom states like $\Lambda_{b}(bud)$, corresponding
cross sections are predicted to be suppressed as $m_{c}^{2}/m_{b}^{2}\sim 1/10$ in comparison with
the case of charm production. Evidence for the forward $\Lambda_{b}$ production in the $pp$
collisions at the ISR energy was reported in Refs.~\cite{ISR-B-1,ISR-B-2}.

Rare seven-quark fluctuations of the type $\left\vert uudc\bar{c}c\bar{c}\right\rangle$ in the
proton wave function can lead to the production of two $J/\psi$ \cite{brod-7-quark} or a
double-charm baryon state at large $x_{F}$ and low $p_{T}$. Double $J/\psi$ events with a high
combined $x_{F}\geq 0.5$ have been detected in the NA3 experiment \cite{NA3-7-quark}. An
observation of the double-charmed baryon $\Xi^{+}_{cc}(3520)$ with mean $\langle x_{F}\rangle\simeq
0.33$ has been reported by the SELEX collaboration at FNAL \cite{SELEX-7-quark}.

\end{document}